\documentclass[twocolumn]{aastex631}
\usepackage{rotating}

\begin{document}


\title{Discovery of a collimated jet from the low luminosity protostar IRAS 16253$-$2429 in a quiescent accretion phase with the JWST}


\author[0000-0002-0554-1151]{Mayank Narang}
\affiliation{Academia Sinica Institute of Astronomy \& Astrophysics, \\ 11F of Astro-Math Bldg., No.1, Sec. 4, Roosevelt Rd., Taipei 10617, Taiwan}
\affiliation{Department of Astronomy and Astrophysics Tata Institute of Fundamental Research \\ Homi Bhabha Road, Colaba, Mumbai 400005, India}
\author[0000-0002-3530-304X]{P. Manoj}
\affiliation{Department of Astronomy and Astrophysics Tata Institute of Fundamental Research \\ Homi Bhabha Road, Colaba, Mumbai 400005, India}

\author[0000-0002-9497-8856]{Himanshu Tyagi}
\affiliation{Department of Astronomy and Astrophysics Tata Institute of Fundamental Research \\ Homi Bhabha Road, Colaba, Mumbai 400005, India}

\author[0000-0001-8302-0530]{Dan M. Watson}
\affiliation{University of Rochester, Rochester, NY, US}

\author[0000-0001-7629-3573]{S. Thomas Megeath}
\affiliation{University of Toledo, Toledo, OH, US}

\author[0000-0002-6136-5578]{Samuel Federman}
\affiliation{University of Toledo, Toledo, OH, US}

\author[0000-0001-8790-9484]{Adam E. Rubinstein}
\affiliation{University of Rochester, Rochester, NY, US}

\author[0000-0002-6447-899X]{Robert Gutermuth}
\affiliation{University of Massachusetts Amherst, Amherst, MA, US}

\author[0000-0001-8876-6614]{Alessio Caratti o Garatti}
\affiliation{INAF-Osservatorio Astronomico di Capodimonte, IT}

\author[0000-0002-1700-090X]{Henrik Beuther}
\affiliation{Max Planck Institute for Astronomy, Heidelberg, Baden Wuerttemberg, DE}

\author[0000-0001-7491-0048]{Tyler L. Bourke}
\affiliation{SKA Observatory, Jodrell Bank, Lower Withington, Macclesfield SK11 9FT, UK}

\author[0000-0001-7591-1907]{Ewine F. Van Dishoeck}
\affiliation{Leiden Observatory, Universiteit Leiden, Leiden, Zuid-Holland, NL}
\affiliation{Max-Planck Institut f\"ur Extraterrestrische Physik, Garching bei München, DE}

\author[0000-0001-5175-1777]{Neal J. Evans II}
\affiliation{Department of Astronomy, The University of Texas at Austin, 2515 Speedway, Stop C1400, Austin, Texas 78712-1205, USA}

\author[0000-0002-7506-5429]{Guillem Anglada}
\affiliation{Instituto de Astrof{\'i}sica de Andaluc{\'i}a, CSIC, Glorieta de la
Astronom{\'i}a s/n, E-18008 Granada, ES}

\author[0000-0002-6737-5267]{Mayra Osorio}
\affiliation{Instituto de Astrof{\'i}sica de Andaluc{\'i}a, CSIC, Glorieta de la
Astronom{\'i}a s/n, E-18008 Granada, ES}

\author[0000-0002-5812-9232]{Thomas Stanke}
\affiliation{Max-Planck Institut f\"ur Extraterrestrische Physik, Garching bei München, DE}

\author[0000-0002-5943-1222]{James Muzerolle}
\affiliation{Space Telescope Science Institute, Baltimore, MD, US}

\author[0000-0002-4540-6587]{Leslie W. Looney}
\affiliation{Department of Astronomy, University of Illinois, 1002 West Green St, Urbana, IL 61801, USA}
\affil{National Radio Astronomy Observatory, 520 Edgemont Rd., Charlottesville, VA 22903 USA} 

\author[0000-0001-8227-2816]{Yao-Lun Yang}
\affiliation{RIKEN Cluster for Pioneering Research, Wako-shi, Saitama, 351-0106, Japan}

\author[0000-0002-6195-0152]{John J. Tobin}
\affil{National Radio Astronomy Observatory, 520 Edgemont Rd., Charlottesville, VA 22903 USA}

\author[0000-0001-9443-0463]{Pamela Klaassen}
\affiliation{United Kingdom Astronomy Technology Centre, Edinburgh, GB}

\author[0000-0003-3682-854X]{Nicole Karnath}
\affiliation{Space Science Institute, Boulder, CO, US}
\affiliation{Center for Astrophysics Harvard \& Smithsonian, Cambridge, MA, US}

\author[0000-0002-4026-126X]{Prabhani Atnagulov}
\affiliation{Ritter Astrophysical Research Center, Dept. of Physics and Astronomy, University of Toledo, Toledo, OH, US}

\author[0000-0001-7826-7934]{Nashanty Brunken}
\affiliation{Leiden Observatory, Universiteit Leiden, Leiden, Zuid-Holland, NL}

\author[0000-0002-3747-2496]{William J. Fischer}
\affiliation{Space Telescope Science Institute, 3700 San Martin Drive, Baltimore, MD 21218, US}

\author[0000-0001-9800-6248]{Elise Furlan}
\affiliation{Caltech/IPAC, Pasadena, CA, US}

\author[0000-0003-1665-5709]{Joel Green}
\affiliation{Space Telescope Science Institute, 3700 San Martin Drive, Baltimore, MD 21218, US}

\author[0000-0002-2667-1676]{Nolan Habel}
\affiliation{Jet Propulsion Laboratory, Pasadena, CA, US}

\author[0000-0003-1430-8519]{Lee Hartmann}
\affiliation{University of Michigan, Ann Arbor, MI, US}

\author[0000-0002-8115-8437]{Hendrik Linz}
\affiliation{Max Planck Institute for Astronomy, Heidelberg, Baden Wuerttemberg, DE}
\affiliation{Friedrich-Schiller-Universit\"at, Jena, Th\"uringen, DE}

\author[0000-0002-4448-3871]{Pooneh Nazari}
\affiliation{Leiden Observatory, Universiteit Leiden, Leiden, Zuid-Holland, NL}

\author[0000-0002-0557-7349]{Riwaj Pokhrel}
\affiliation{Ritter Astrophysical Research Center, Dept. of Physics and Astronomy, University of Toledo, Toledo, OH, US, 43606 }

\author[0000-0002-5350-0282]{Rohan Rahatgaonkar}
\affiliation{Gemini South Observatory, La Serena, CL}

\author[0000-0001-6144-4113]{Will R. M. Rocha}
\affiliation{Laboratory for Astrophysics, Leiden Observatory, Universiteit Leiden, Leiden, Zuid-Holland, NL}

\author[0000-0002-9209-8708]{Patrick Sheehan}
\affil{National Radio Astronomy Observatory, 520 Edgemont Rd., Charlottesville, VA 22903 USA} 

\author[0000-0002-7433-1035]{Katerina Slavicinska}
\affiliation{Leiden Observatory, Universiteit Leiden, Leiden, Zuid-Holland, NL}

\author[0000-0003-2300-8200]{Amelia M.\ Stutz}
\affiliation{Departamento de Astronom\'{i}a, Universidad de Concepci\'{o}n,Casilla 160-C, Concepci\'{o}n, Chile}

\author[0000-0002-9470-2358]{Lukasz Tychoniec}
\affiliation{European Southern Observatory,
Garching bei M\"unchen, DE}

\author[0000-0002-0826-9261]{Scott Wolk}
\affiliation{Center for Astrophysics Harvard \& Smithsonian, Cambridge, MA, US}

\begin{abstract}
Investigating Protostellar Accretion (IPA) is a JWST Cycle~1 GO program that uses NIRSpec IFU and MIRI MRS to obtain 2.9--28~$\mu$m spectral cubes of young, deeply embedded protostars with luminosities of 0.2 to 10,000~L$_{\odot}$ and central masses of 0.15 to 12~M$_{\odot}$. 
In this Letter, we report the discovery of a highly collimated atomic jet from the Class~0 protostar IRAS~16253$-$2429, the lowest luminosity source ($L_\mathrm{bol}$ = 0.2 $L_\odot$) in the IPA program. The collimated jet is detected in multiple [Fe~II] lines, [Ne~II], [Ni~II], and H~I lines, but not in molecular emission. The atomic jet has a velocity of about 169~$\pm$~15~km\,s$^{-1}$, after correcting for inclination. The width of the jet increases with distance from the central protostar from 23 to~60 au,  corresponding to an opening angle of 2.6~$\pm$~0.5\arcdeg. By comparing the measured flux ratios of various fine structure lines to those predicted by simple shock models, we derive a shock {speed} of 54~km\,s$^{-1}$ and a preshock density of 2.0$\times10^{3}$~cm$^{-3}$ at the base of the jet.  {From these quantities and using a suite of jet models and extinction laws we compute a mass loss rate between $0.4 -1.1\times10^{-10}~M_{\odot}$~yr~$^{-1}$.} The low mass loss rate is consistent with simultaneous measurements of low mass accretion rate ($2.4~\pm~0.8~\times~10^{-9}~M_{\odot}$~yr$^{-1}$) for IRAS~16253$-$2429 from  JWST observations (Watson et al. in prep), indicating that the protostar is in a quiescent accretion phase. Our results demonstrate that very low-mass protostars can drive highly collimated, atomic jets, even during the quiescent phase.

\end{abstract}

\section{Introduction} \label{sec:intro}

Jets and outflows from protostars and pre-main sequence stars play a crucial role in the process of star formation.  These dynamic phenomena are instrumental in facilitating the removal of angular momentum from the surrounding disk material, thereby enabling efficient accretion of mass onto the central object \citep[e.g.,][]{2007prpl.conf..261S,2007prpl.conf..277P,2016ARA&A..54..135H,2014prpl.conf..451F,B16,2021NewAR..9301615R}.  An accretion-driven outflows/jets also play vital roles in shaping the Initial Mass Function (IMF) \citep[e.g.,][]{2022MNRAS.515.4929G,2023arXiv230905397L} and can inject energy and momentum into the surroundings, and disperse a significant fraction of the protostellar envelope, which limits star formation efficiency  \citep[e.g.,][]{2010ApJ...710L.142F,2014prpl.conf..451F,2015MNRAS.450.4035F,2021ApJ...912L..19P,2021ApJ...911..153H,2023ApJ...954...93A,2023ApJ...947...25H}.

\begin{figure*}
\centering
\includegraphics[width=1\linewidth]{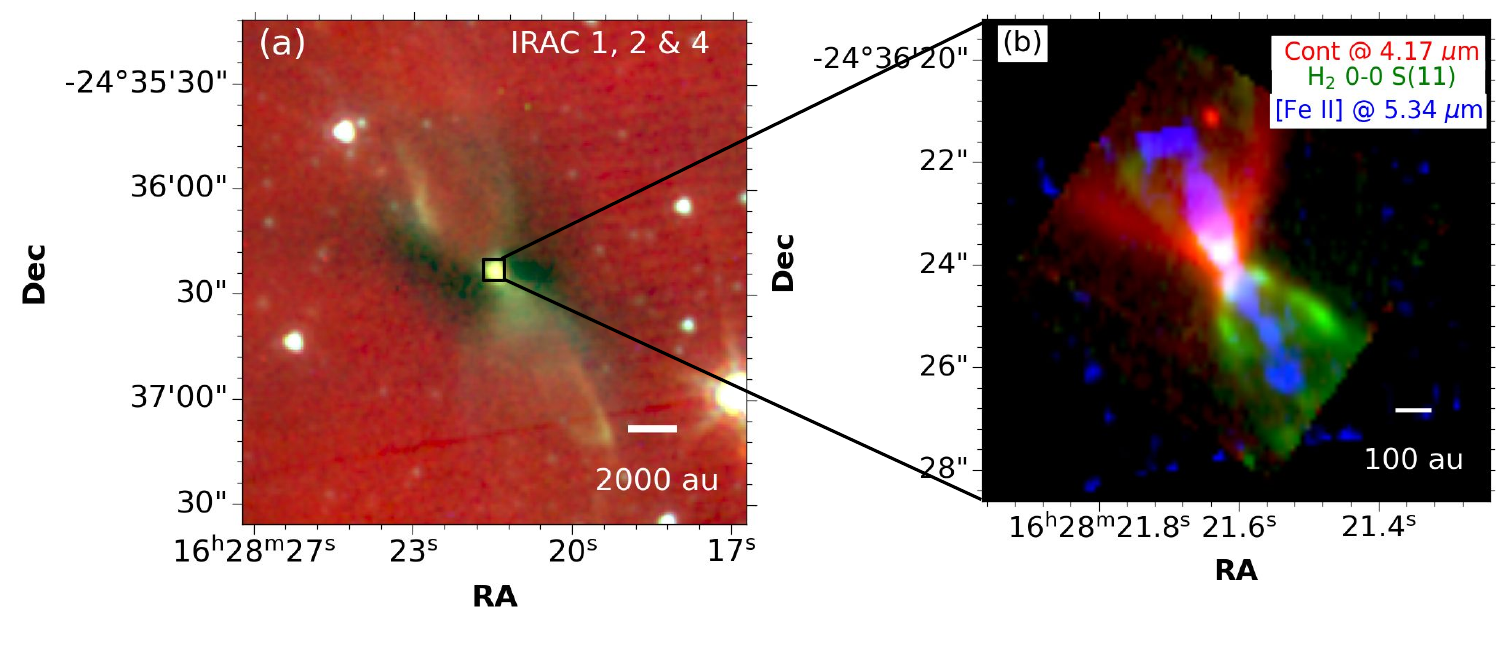}
\caption{(a) The Spitzer IRAC~three color (IRAC~3.6, 4.5 and 8 $\mu$m  are red, green and blue, respectively) image of the IRAS 16253$-$2429 field. The field of view is 2\arcmin $\times$ 2\arcmin. The black square in the center represents {a 6\arcsec $\times$ 6\arcsec{} field centered on the JWST observations.}  (b) A zoomed-in view of the NIRSpec and MIRI channel-1 Short (A) field with the red representing continuum emission at 4.17~$\mu$m, the green showing H$_2$ 0-0 S(11) at 4.18~$\mu$m and the blue depicting [Fe~II] at 5.34~$\mu$m. A local continuum has been subtracted to make the line maps.  }
\label{Fig1}
\end{figure*}

Despite their unquestionable significance, the precise origins of protostellar jets and outflows remain uncertain and continue to be a subject of debate. Among the various proposed mechanisms, the most widely accepted one is magnetocentrifugal acceleration  \citep{1982MNRAS.199..883B,1983ApJ...274..677P,1992ApJ...394..117P, 1993ApJ...410..218W, 1994ApJ...429..808N,1994ApJ...429..781S, 2005ApJ...632L.135M,2008ApJ...678.1109M,2014prpl.conf..451F}. According to this model, material situated near the surface of the inner parts of the disk is ejected centrifugally along the rotating magnetic field lines. As the outflow progresses beyond the Alfv\'en surface where the kinetic energy matches the magnetic energy, the dynamics begin to dominate over the magnetic field, resulting in a predominantly toroidal magnetic configuration. The tension force generated by these magnetic loops focuses the outflowing material into a narrow jet along the rotational axis of the circumstellar disk that originally launched the (MHD) wind \citep{1982MNRAS.199..883B,1996ASIC..477..249S,2021NewAR..9301615R}. As a result, highly collimated bipolar jets are formed. 


\begin{figure*}
\centering
\includegraphics[width=0.85\linewidth]{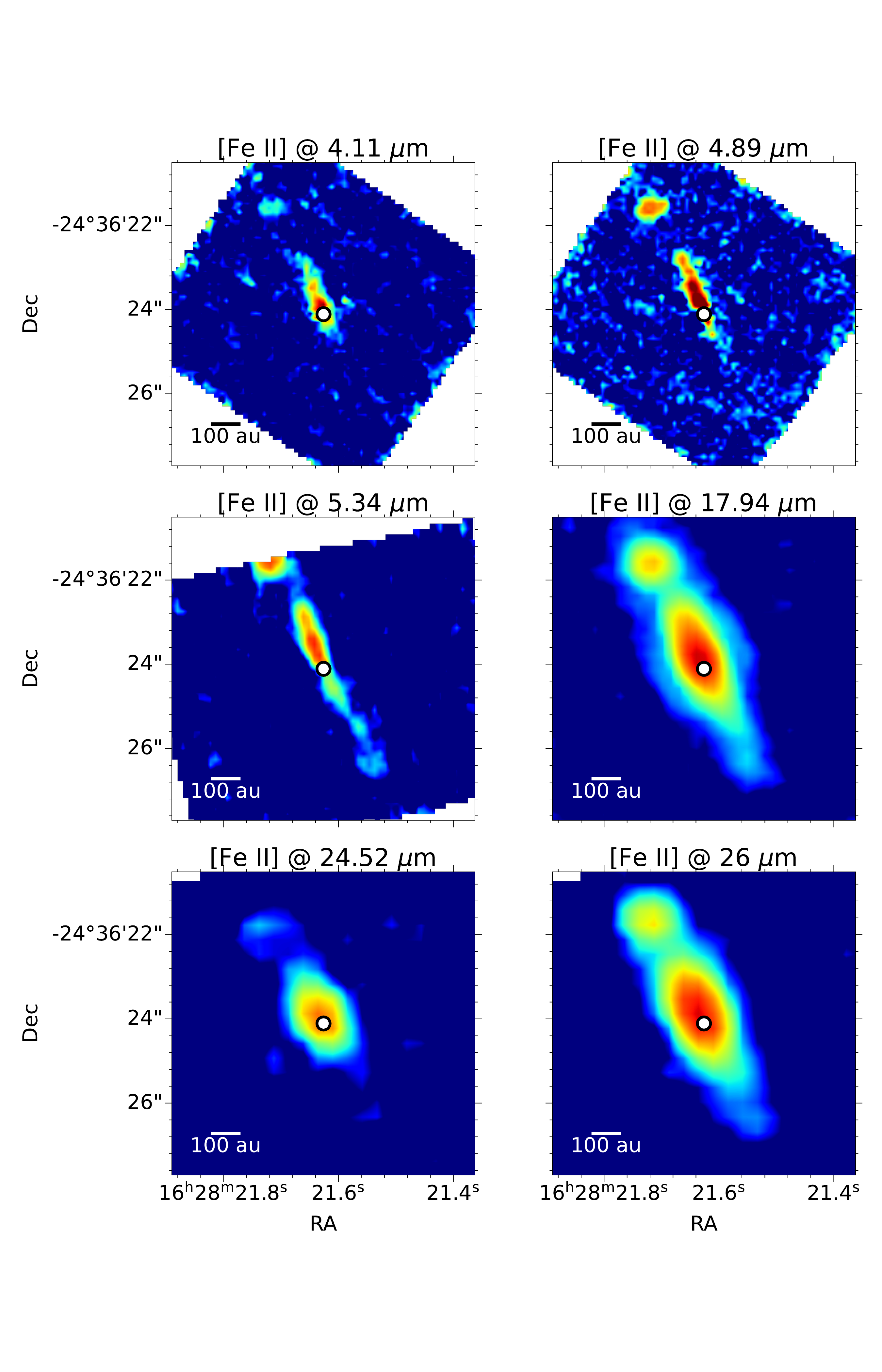}
\caption{The morphology of [Fe~II] lines detected towards IRAS 16253$-$2429. All these lines trace the collimated jet launched from the protostar.   All images are shown on the same spatial scale.  The white solid circle is the MIRI MRS 14~$\mu$m continuum position. The scale bar corresponding to 100~au is shown in the bottom left corner. }
\label{Fig2}
\end{figure*}

Infrared emission lines of [Fe~II] and H$_2$ are well-known tracers of shocked gas and have been used to trace jets and outflows from protostars \citep[e.g.][]{2006A&A...449.1077C,2014A&A...564A..11A, 2019AJ....158..107R,2016ApJ...828...52W, B16, 2022ApJ...941L..13Y, 2023A&A...673A.121B, 2023ApJ...951L..32H,2023ApJ...948...39R, Ray23}. These lines serve as excellent diagnostics for deriving the physical conditions of the collimated jets as well as the shocked molecular gas in the outflow \citep{Ray23}. Harnessing the spatial and spectral resolution as well as the sensitivity of JWST, we can gain a deeper understanding of the launching of and feedback from the protostellar jets/outflows. 

\begin{table*}[]
\centering
\caption{{The log of the JWST observations (Program ID 1802) presented in this paper }} 

\begin{tabular}{|l|c|c|c|c|c|c|}
\hline
Obs & Obs No& RA & Dec & Hours & Start UT & End UT \\\hline
NIRSpec IFU IRAS 16253$-$2429& 015 & 16 28 21.62$^*$ & -24 36 24.16$^*$ & 3.43 & Jul 22, 2022 15:09:15  & Jul 22, 2022 17:59:43 \\
MIRI MRS IRAS 16253$-$2429 & 013& 16 28 21.62$^*$ & -24 36 24.16$^*$ & 7.43 & Jul 23, 2022 04:04:41 & Jul 23, 2022 09:58:57 \\
MIRI MRS IRAS 16253$-$2429& 014& 16 28 27.756 & -24 37 46.95 & 1.86 & Jul 23, 2022 10:02:33 & Jul 23, 2022 11:35:19\\ 
 background & &  & & & & \\ \hline
\end{tabular}
\label{log}
\tablecomments{{$^*$ The coordinates are the field center of the 2 $\times$ 2 mosaics.}}
\end{table*}

Investigating Protostellar Accretion (IPA) across the mass spectrum is a JWST Cycle~1 medium GO program (Program ID 1802, PI Tom Megeath; \citealt{2021jwst.prop.1802M},  \citealt{2023arXiv231003803F}, \citep{2023arXiv231207807R} , Watson et al., prep). The IPA program aims to study the processes that drive protostellar accretion, and accretion-driven jets/outflows as a function of protostellar mass. 
The IPA protostellar sample consists of five sources that range in bolometric luminosities from 0.2 to 10,000~$L_{\odot}$ and in central masses from 0.15 to 12~$M_{\odot}$. These protostars are in the Class~0 phase, during which most of the stellar mass is accreted \citep[e.g.,][]{2017ApJ...840...69F,2023ApJ...944...49F, 2023arXiv230812689N}.

The protostar IRAS~16253$-$2429 is the lowest luminosity source in the IPA sample. It is located in a relatively isolated region of the Ophiuchus molecular cloud \citep{2004A&A...426..171K,2006A&A...447..609S, 2010ApJ...720...64B}. The protostar has a bolometric luminosity of only 0.2~$L_\odot$ and a bolometric temperature of 42~K (Pokhrel et al., in preparation; also see \citealt{2023ApJS..266...32P}) and an internal luminosity below 0.15~$L_\odot$ (Pokhrel et al., in preparation). Being only at a distance of 140~pc (average of $\rho$-Ophiuchi molecular cloud members \citealt{2020A&A...633A..51Z}) and relatively isolated,  IRAS 16253$-$2429 is an ideal target to study the morphology and kinematics of jets from very low mass ($<0.2$~$M_\odot$) protostars \citep{ yen17,2023ApJ...954..101A}. 

Extensive studies have been conducted on this protostar, particularly at (sub)millimeter wavelengths \citep{2006A&A...447..609S, 2011ApJ...740...45T, yen17,2019ApJ...871..100H}. These observations have revealed the presence of a central protostar associated with a bipolar outflow identified in CO, as well as a Keplerian disk. {However, no collimated molecular jet in the low-J lines of CO or SiO has been detected from the system \citep{2023ApJ...954..101A}. This is consistent with the low detection rate of molecular jets from low luminosity protostars \citep[e.g.,][]{2021A&A...648A..45P}.} Recent observations with ALMA have estimated the dynamical mass of the central source to be between 0.12-0.17~$M_\odot$ \citep{2023ApJ...954..101A}, {indicating that IRAS 16253$-$2429  has already accreted a  mass above the hydrogen-burning limit. The protostar also has a dust disk with a diameter of 15~au \citep{2023ApJ...954..101A} with a mass of  $2 \times 10^{-3}\,M_\odot$.  The envelope mass of IRAS~16253$-$2429 has been estimated from (sub)mm observations to be in the range of 0.2 to 1~$M_\odot$ (within a radius of $\sim$ 4200 au), \citep{2006A&A...447..609S, 2008ApJ...684.1240E, 2011ApJ...740...45T}. The disk-to-envelope mass ratio for IRAS 16253$-$2429, therefore, is $\ll$~0.01 and is consistent with the values reported for Class~0 protostars by \cite{2023ApJ...944...49F}. }

Figure~\ref{Fig1} shows the Spitzer Infrared Array Camera (IRAC) images of the IRAS~16253$-$2429 field, which reveal a bipolar hourglass structure in the Northeast to Southwest direction. This distinctive shape is traced by the outflow cavities in H$_2$ and the continuum. The bipolar outflow from the source extends up to 2\arcmin\ in size, corresponding to a physical scale of 16800~au or 0.08~pc. In addition to the imaging, the Spitzer IRS instrument also carried out spectral scan mapping of the protostar \citep{2010ApJ...720...64B}. The Spitzer IRS spectra detected six pure rotational H$_2$ (0-0 S(2) to 0-0 S(7)) lines along the protostellar outflow. However, no fine structure lines were detected from the protostar with Spitzer/IRS.

\section{Observations and Data reduction}

Our primary objective with the JWST observations was to perform spectral imaging by combining the Integral Field Unit (IFU) observations from the Mid-Infrared Instrument (MIRI) Medium Resolution Spectrograph (MRS) \citep{2015PASP..127..584R,2015PASP..127..595W} and the Near Infrared Spectrograph (NIRSpec) \citep{2022A&A...661A..80J, 2022A&A...661A..82B} covering 2.87 to 28~$\mu$m. The NIRSpec IFU and MIRI MRS observations were conducted nearly simultaneously. The details of the observation log are provided in Table \ref{log}. 

{For the NIRSpec IFU observations, we utilized the F290L/G395M filter/grating combination which has a wavelength coverage from 2.87–5.10 $\mu$m and a nominal resolving power of $\sim 1000$\footnote{https://jwst-docs.stsci.edu/jwst-near-infrared-spectrograph/nirspec-instrumentation/nirspec-dispersers-and-filters}.} To ensure comprehensive {coverage of the protostellar jet/outflow}, we employed a $2\times2$ mosaic pattern with a 10\% overlap and a 4-point dither mode. The NIRSpec {mosaic covers $\sim 6$\arcsec{}$\times 6$\arcsec{} field centered at RA =  16$^h$28$^m$21.62$^s$  and Dec =-24$^d$36$^m$24.16$^s$}. The total exposure time, which accounted for overheads, amounted to 3.43~hours (for further detail, see \citealt{2023arXiv231003803F}). {Using the background star in the NIRSpec field of IRAS 16253-2429, \cite{2023arXiv231003803F} measured the NIRSpec FWHM to be between 0.\arcsec19~$-$~0.\arcsec20.}

\begin{figure*}
\centering
 \includegraphics[width=0.85\linewidth]{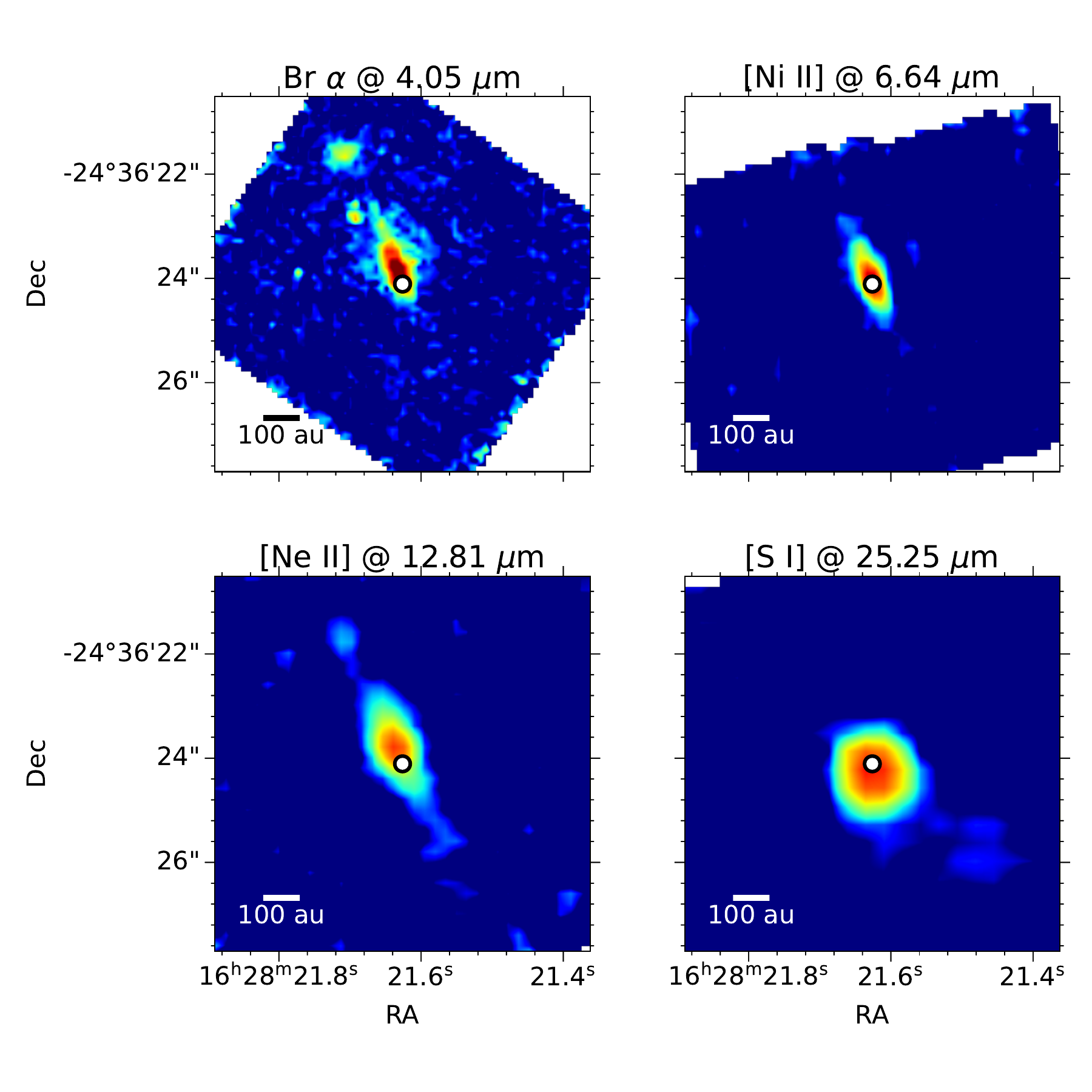}
\caption{Morphology of the Br-$\alpha$ line and the other atomic fine-structure lines detected towards IRAS 16253$-$2429.  All images are shown on the same spatial scale.  The white solid circle is the MIRI MRS 14~$\mu$m continuum position. The scale bar corresponding to 100~au is shown in the bottom right corner. }
\label{Fig3}
\end{figure*}

{IRAS 16253$-$2429 was observed using all four channels of the MIRI MRS with wavelength coverage from 4.9 to 27.9 $\mu$m, with a spectral resolving power ranging from $\sim$1500 - 4000 \citep{2023MNRAS.523.2519J}\footnote{ also see https://jwst-docs.stsci.edu/jwst-mid-infrared-instrument/miri-observing-modes/miri-medium-resolution-spectroscopy}. The spatial resolution of MIRI MRS ranges from $\sim$ 0.\arcsec27 to 1\arcsec \citep{2023AJ....166...45L}.} Similar to the NIRSpec observations, we employed a $2\times2$ mosaic pattern with a 10\% overlap and a 4-point dither for the MIRI observations. The total exposure time, including overheads, for the science target amounted to 7.43~hours. Additionally, a dedicated background observation with a similar setup and an exposure time of 1.86~hours (including overheads) was obtained. The dedicated background field was positioned approximately 118\arcsec{} away from the target.  The {spatial coverage} for MIRI varies from channel to channel with channel-1 short {covering}  $\sim$6\arcsec $\times$ 6\arcsec while for channel-4 long {the coverage is} $\sim$15\arcsec $\times$ 15\arcsec. {The orientation and the field coverage of the MIRI MRS instrument is also different from the NIRSpec IFU \footnote{https://jwst-docs.stsci.edu/jwst-observatory-characteristics/jwst-position-angles-ranges-and-offsets}. In addition, MIRI parallel imaging observations were obtained at both MIRI pointings.}

We reduced the NIRSpec IFU using the JWST pipeline version {1.9.5} and the JWST Calibration References Data System context version {jwst\_1069.pmap}. Our analysis of the NIRSpec data showed the presence of several hot pixels that were being missed by the JWST outlier detection step. As these hot pixels were a function of wavelength and time, a universal dark mask could not be used. To remedy this, we used a custom outlier detection algorithm for the NIRSpec observations to remove the hot pixels (see \citealt{2023arXiv231003803F}).  

\startlongtable 
\begin{deluxetable*}{cccccccc}
    \tablewidth{0pt}
    \tablecaption{The ionic and atomic lines reported in this work.}
    \tablehead{
    \colhead{Wavelength} & \multicolumn{4}{c}{Species and transition}  &  \colhead{$A_{ul}$} & \colhead{$E_{up}$} & \colhead{ref}  \\
    \colhead{$\micron$} & \multicolumn{4}{c}{Upper state - Lower  state}  &  \colhead{(s$^{-1}$)}   &  \colhead{(K)} & \colhead{}  \\
    }
    \decimals
    \startdata
4.052 	 & 	H-I (Br$\alpha$)	& 	$n = 5-4$	 & 		 & 		 & 	 2.69 $\times 10^{6}$ 	 & 		151492 & 2	 \\
4.115	 & 	[Fe II]	 & 	${}^4 F_{7/2}-{}^6 D_{9/2}$	 & 		 & 		 & 	 7.71 $\times 10^{-6}$ 	 & 		3496 & 1,2	 \\
4.889	 & 	[Fe II]	 & 	${}^4 F_{7/2}-{}^6 D_{7/2}$	 & 		 & 		 & 	 9.3 $\times 10^{-5}$ 	 & 		3496	& 2		 \\
5.340	 & 	[Fe II]	 & 	${}^4 F_{9/2}-{}^6 D_{9/2}$	 & 		 & 		 & 	 2.37 $\times 10^{-5}$ 	 & 		2694 &	1,2	 \\
12.814	 & 	[Ne II]	 & 	${}^2 P_{1/2}-{}^2 P_{3/2}$	 & 		 & 		 & 	 8.32 $\times 10^{-3}$ 	 & 		1123 &	2	 \\
17.936	 & 	[Fe II]	 & 	${}^4 F_{7/2}-{}^4 F_{9/2}$	 & 		 & 		 & 	 5.84 $\times  10^{-3}$	 & 		3496	& 2		 \\
24.519	 & 	[Fe II]	 & 	${}^4 F_{5/2}-{}^4 F_{7/2}$	 & 		 & 		 & 	 3.93 $\times 10^{-3}$ 	 & 		4083 & 2			 \\
25.249	 & 	[S I]	 & 	${}^3 P_{1}-{}^3 P_{2}$	 & 		 & 		 & 	 1.4 $\times 10^{-3}$ 	 & 		570	&	 2 \\
25.988	 & 	[Fe II]	 & 	${}^6 D_{7/2}-{}^6 D_{9/2}$	 & 		 & 		 & 	 2.14 $\times 10^{-3}$ 	 & 		554	&	 1,2 \\ \enddata
    \tablecomments{(1) \cite{2018PhRvA..98a2706T}; (2)  \cite{NIST}}
     \label{Table3}
      \vspace{-0.3in}
\end{deluxetable*}

\begin{figure*}
\centering
\includegraphics[width=0.45\linewidth]{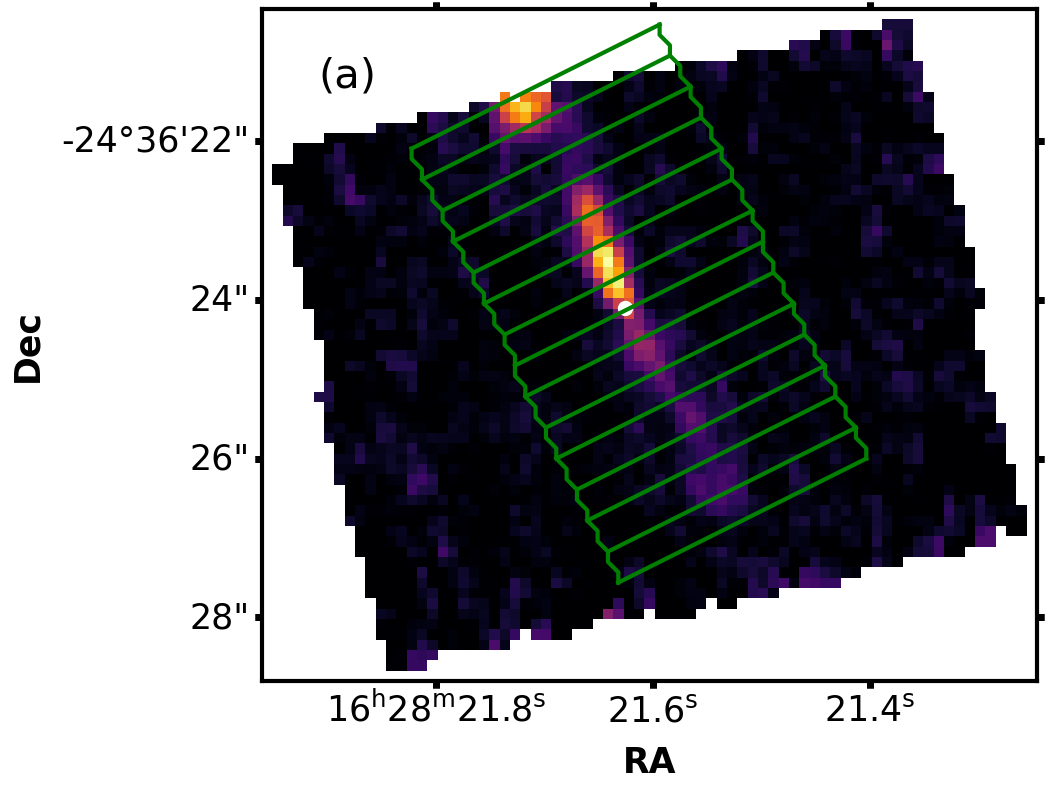}\includegraphics[width=0.35\linewidth]{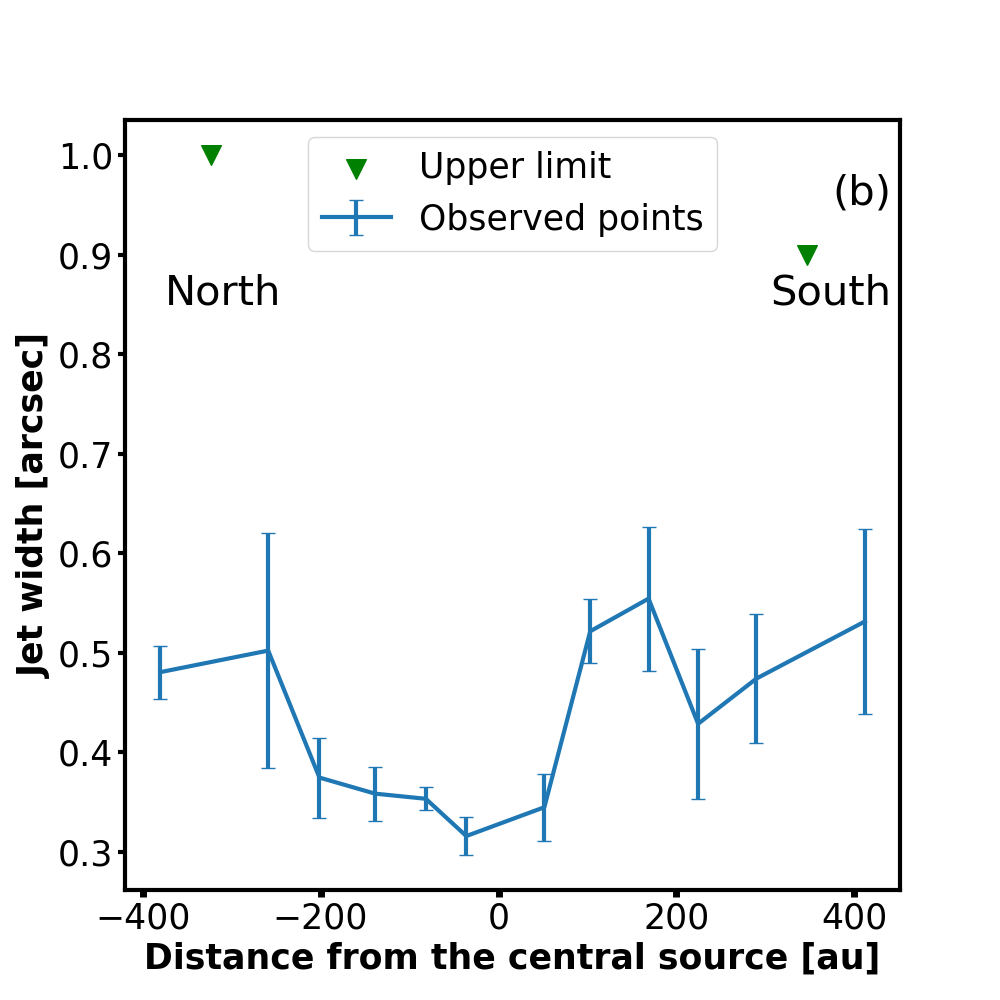}
\includegraphics[width=0.35\linewidth]{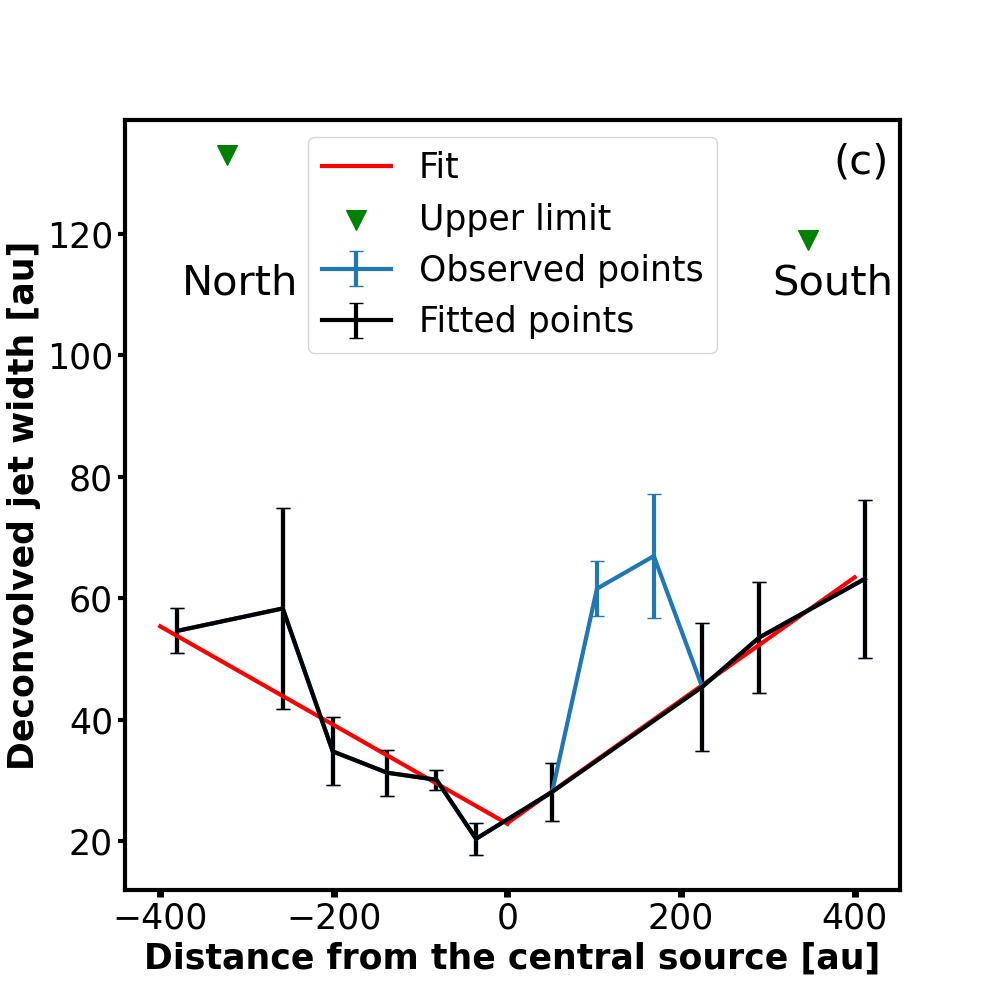}
\caption{(a) The horizontal slices in green overlaid on top of the jet traced in [Fe~II] at 5.34~$\mu$m mark slices {that are 3 pixels wide along the jet direction. The jet width is computed from each of these green slices (see Section 3.2)} The white circle represents the protostellar position.  (b) The width (FWHM) of the jet as a function of distance from the central protostar. The negative distance corresponds to the northern {jet}, and the positive distance corresponds to the southern {counter-jet}. (c) The deconvolved width (FWHM) of the jet as a function of distance from the central protostar. The fit to the jet width as a function of distance is shown as the red solid line. The green triangles in sub-figure (b) and (c) show the upper limits to the width at those points. }
\label{Fig4}
\end{figure*}

\begin{figure*}
\centering
 \includegraphics[width=0.8\linewidth]{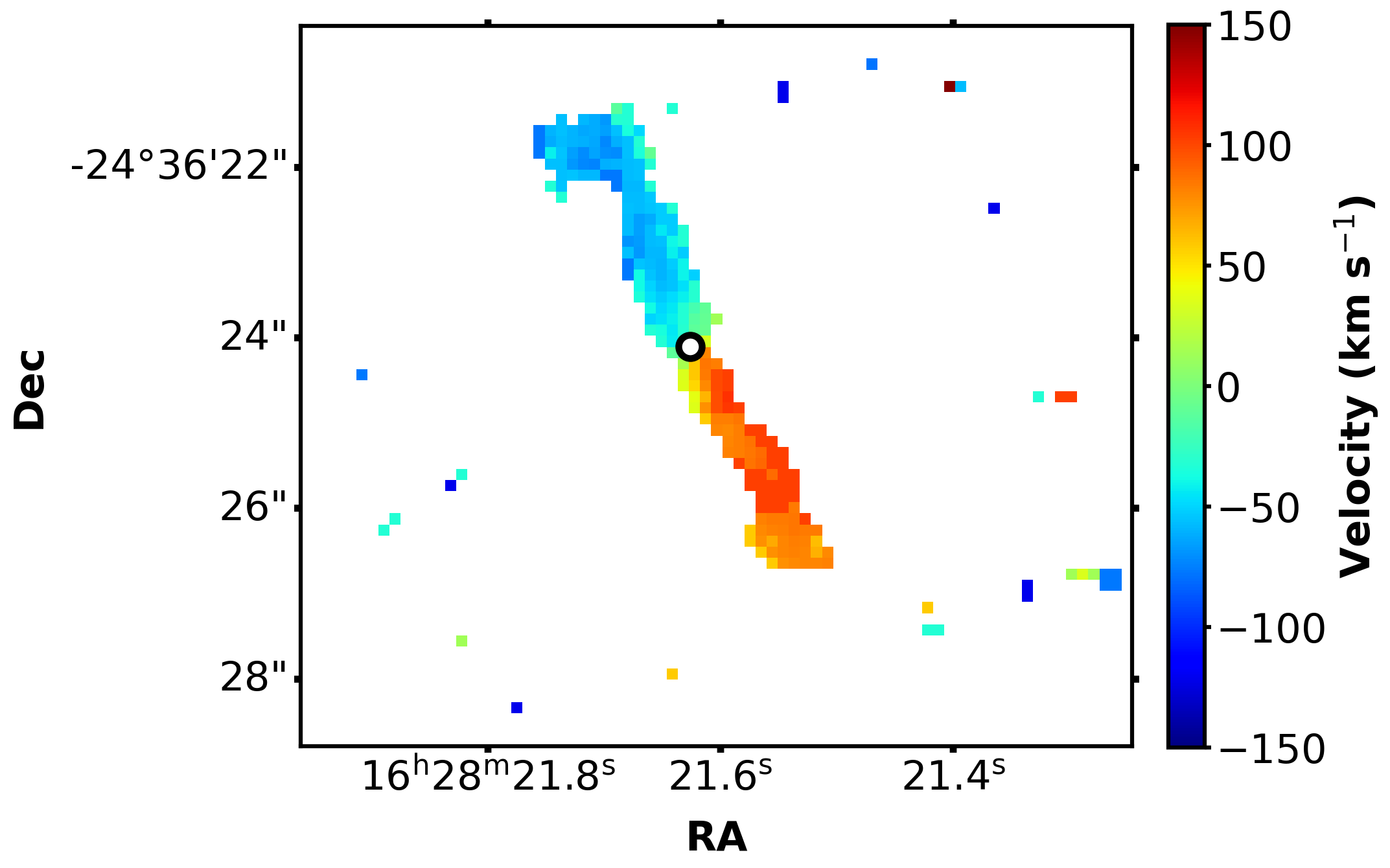}
\caption{The moment 1 map for the  [Fe~II] line at 5.34~$\mu$m.  The white solid circle is the MIRI MRS 14~$\mu$m continuum position.}
\label{Fig6}
\end{figure*}

To reduce the MIRI data we used the JWST pipeline version {1.11.3} and the JWST Calibration References Data System context version {jwst\_1100.pmap}. We used the standard Stage-1 JWST pipeline \textit{Detector1Pipeline} to reduce the MIRI MRS data starting from \textit{uncal} data. In Stage-2 (\textit{Spec2Pipeline}), the dedicated background was subtracted,  and the residual fringe correction was also performed. The Stage-3 \textit{Spec3Pipeline} step was then run with the \textit{CubeBuildStep}  set to \textit{'band'} mode such that each channel and each band are reduced as a separate fits file. However, on examining the dedicated background IFU image, we detected emission from the  H$_2$ S (1) (17.035~$\mu$m) and H$_2$ S(2) (12.279~$\mu$m) molecular lines. Therefore we also reduced the MIRI MRS data without subtracting the dedicated background. We have used these non-background subtracted data cubes for measuring the extinction in Section 3.3.  

During our analysis of the data obtained from the MIRI MRS and NIRSpec IFU instruments, we found a positional offset between the two instruments. To accurately determine and quantify the offset between the two instruments, we compared the overlapping wavelength regions  of MIRI MRS channel 1-short and the NIRSpec IFU cubes. The positional offset between these two instruments (MIRI - NIRSpec) was $\Delta$RA = 0\arcsec.37 and $\Delta$Dec = 0\arcsec.67. We further refined the position of the MIRI MRS field by aligning it with Spitzer and Gaia data of the stellar sources detected in the MIRI parallel image.  We then adjusted the NIRSpec position using the determined offset between it and the MRS.  We use these refined positions in our analysis. For further elaboration and in-depth information concerning this positional offset, see \cite{2023arXiv231003803F}.

\begin{figure*}
\centering
\includegraphics[width=0.8\linewidth]{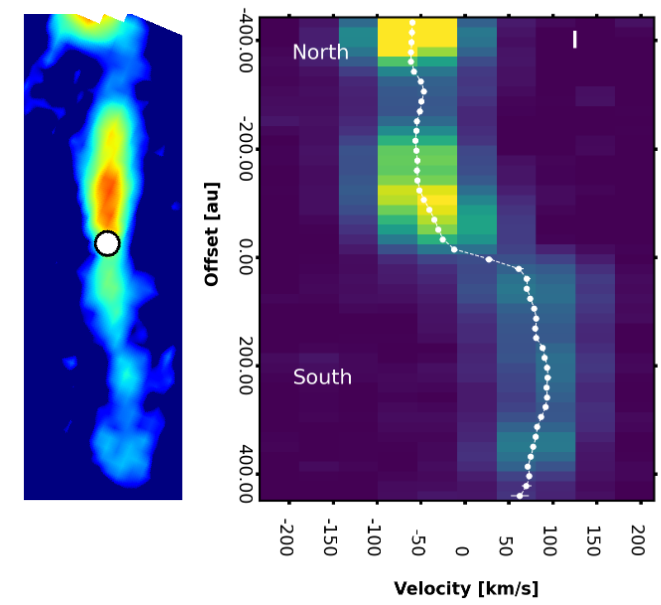}
\caption{  Left: image showing the line map of the [Fe II] line at 5.34 $\mu$m  for comparison to the PV diagram. The width of the image gives the range over which we averaged the jet emission. The circle shows the 14 $\mu$m MIRI continuum position of the protostar.  Right: the PV diagram for the [Fe~II] line at 5.34~$\mu$m. The white circles represent the center of the Gaussian fit to the velocity channels, while the error bars represent the error in determining the center of the Gaussian fit.  The white bar in the top left corner is the  FWHM of the JWST MIRI PSF at 5.34 $\mu$m (0\arcsec.28 or 39.2~au). }
\label{Fig7}
\end{figure*}

\section{Jets from IRAS 16253-2429}
\subsection{Morphology}

The JWST observations of IRAS 16253-2429 with the NIRSpec IFU and MIRI MRS instruments reveal atomic jets, molecular outflows, and envelope cavity structures at spatial scales down to 30~au. In Figure \ref{Fig1}, we present a comparison between the Spitzer IRAC (IRAC~ 3.6, 4.5 and 8~$\mu$m; Figure \ref{Fig1}a) observations of the protostar, covering a field of view of 2\arcmin $\times$ 2\arcmin, and the NIRSpec IFU + MIRI MRS observations (Figure \ref{Fig1}b) focused within the inner 6\arcsec \citep[also see][]{2023arXiv231003803F}. Figure \ref{Fig1}(a) shows the full spatial extent of the protostellar outflow from IRAS 16253$-$2429 as seen by Spitzer. The black rectangle shows the field of view of the JWST observations. Figure \ref{Fig1}(b) shows the zoomed-in inner $\sim$6\arcsec{} region of the protostar as observed with JWST. The 4.17~$\mu$m continuum emission is primarily due to the scattered light from the central source and is shown in red; it traces the extent of the outflow cavity carved out of the envelope.  The molecular outflow traced by H$_2$ 0-0 S(11) emission at 4.18~$\mu$m is shown in green. The atomic jet seen in the [Fe~II] line at 5.34~$\mu$m is shown in blue. The JWST observations show that IRAS~16253$-$2429 exhibits the distinctive ``wasp-waist'' shape \citep{2010ApJ...720...64B}, even within the inner 6\arcsec{} region surrounding the protostar. Figure \ref{Fig1}(b) further shows that the scattered light cavity is much broader than the molecular emission from H$_2$ (also see \citealt{2023arXiv231003803F} and Narang et al. in prep).

The wavelength range covered by the MIRI MRS and NIRSpec IFU instruments encompasses several atomic and fine structure lines, including multiple lines of [Fe~II]. {In Table \ref{Table3}, we have listed the bright atomic and ionic lines detected towards IRAS 16253-2429, along with their wavelength, Einstein~$A$~coefficient $A_{ul}$, and upper state energy $E_{up}$.}

In Figure \ref{Fig2}, we present the line maps of all the detected [Fe~II] lines observed towards IRAS 16253$-$2429. These [Fe~II] lines trace a highly collimated jet originating from the central protostar. This is the first time (at any wavelength) that a collimated jet has been detected from IRAS 16253$-$2429. The jet appears brighter on the northern side and weaker to the south; this is due to the relatively higher line-of-sight extinction toward the southern cavity (see Section 3.4).

We further find that there are knots at both ends of the jet (see Figure \ref{Fig2}) at around 400~au from the central protostar (see also \citealt{2023arXiv231003803F}). The southern knot is best detected in the [Fe~II] line at 5.34~$\mu$m, with a tentative detection in the [Fe~II] lines at 17.936~$\mu$m and 25.988~$\mu$m.  The northern knot is clearly seen in five of the six [Fe~II] line maps presented in Figure \ref{Fig2}, although it is not fully mapped in the 5.34~$\mu$m line, and is only faintly seen in the 24.5~$\mu$m line. If we draw a line from the central protostar to the northern- and southern-most knots we find that the jet seen in [Fe~II] at 5.34~$\mu$m does not fall on this line. The northern knot is clearly offset to the east from the axis of the rest of the northern jet. We also find that the jet has a slight curvature, suggesting that it might be precessing (see also \citealt{2010ApJ...720...64B}).

In Figure \ref{Fig3}, we present maps of other atomic fine-structure (FS) lines along with the Br-$\alpha$ line at 4.052~$\mu$m. We find that the FS lines of [Ne~II] at 12.81~$\mu$m and the [Ni~II]  at 6.64~$\mu$m also trace the jet seen in [Fe~II]. The [Ne~II] jet also shows faint knots at both ends similar to the [Fe~II] jets. The [S~I] emission at 25.25~$\mu$m, however, is concentrated on the protostar and does not appear to trace the jet (see Figure \ref{Fig3}). The Br-$\alpha$ line observed with NIRSpec appears to trace the jet and shows the knot on the northern side (see Figure \ref{Fig3}). Interestingly, Br-$\alpha$ emission is stronger in the extended jets than at the central source. This suggests that the H~I 
lines observed from protostars via spatially unresolved spectroscopy might have a large contribution from the jet component; the dominant jet component makes it difficult to isolate emission in this line from accretion flows \citep[see][]{2016ARA&A..54..135H,2023arXiv231003803F}. 

\subsection{Jet width}
Given the high signal-to-noise and high angular resolution at 5.34~$\mu$m (0.\arcsec28, \citealt{2023AJ....166...45L}), it is possible to measure the width of the jet as a function of distance from the protostar. The jet has a position angle of 23\arcdeg\ (measured from North towards East).  We started by taking slices aligned perpendicular to the jet and averaged them over 3~pixels length along the jet as shown in Figure \ref{Fig4}(a). We then fit the jet cross section in each of these averaged slices with a Gaussian and the FWHM of the Gaussian is taken as the width of the jet in that slice. 

In Figure~\ref{Fig4}(b) we show the width of the jet (FWHM) as a function of distance from the protostar. As can be seen from Figure~\ref{Fig4}(b), the jet appears to expand as we move away from the central protostar. We measure the FWHM of the slices adjacent to the protostar on either side and find the width of the jet close to the central protostar to be 0\arcsec.33~$\pm$~0\arcsec.03. This is comparable to the average PSF FWHM of the MIRI MRS at 5.34~$\mu$m of 0.\arcsec28 \citep{2023AJ....166...45L}. 
 
In Figure~\ref{Fig4}(c) we show the deconvolved jet widths (obtained by subtracting out the MIRI MRS point spread function FWHM in quadrature) as a function of distance from the central protostar. From a linear fit to the jet width as a function of distance from the central protostar, we obtain the opening angle $\Theta$ ($ = {\rm slope /2}$) -- {the half angle subtended by the jet width at the launch point} --  as well as the jet width at the central position (zero intercept). The average slope of the fit is $0.09 \pm 0.02$, corresponding to an opening angle of $2.\arcdeg6 \pm 0.\arcdeg5$. The jet width extrapolated to the source position is $23 \pm 4$~au, which we take as an upper limit for the jet width at the source location. \cite{2023arXiv231003803F} used the NIRSpec observation of the [Fe~II] line at 4.88~$\mu$m to measure the width of the jet from IRAS 16253$-$2429 at its brightest position and found a deconvolved jet width of 0\arcsec.144, which corresponds to 20.2~au. 

\begin{figure*}
\centering
 \includegraphics[width=1\linewidth]{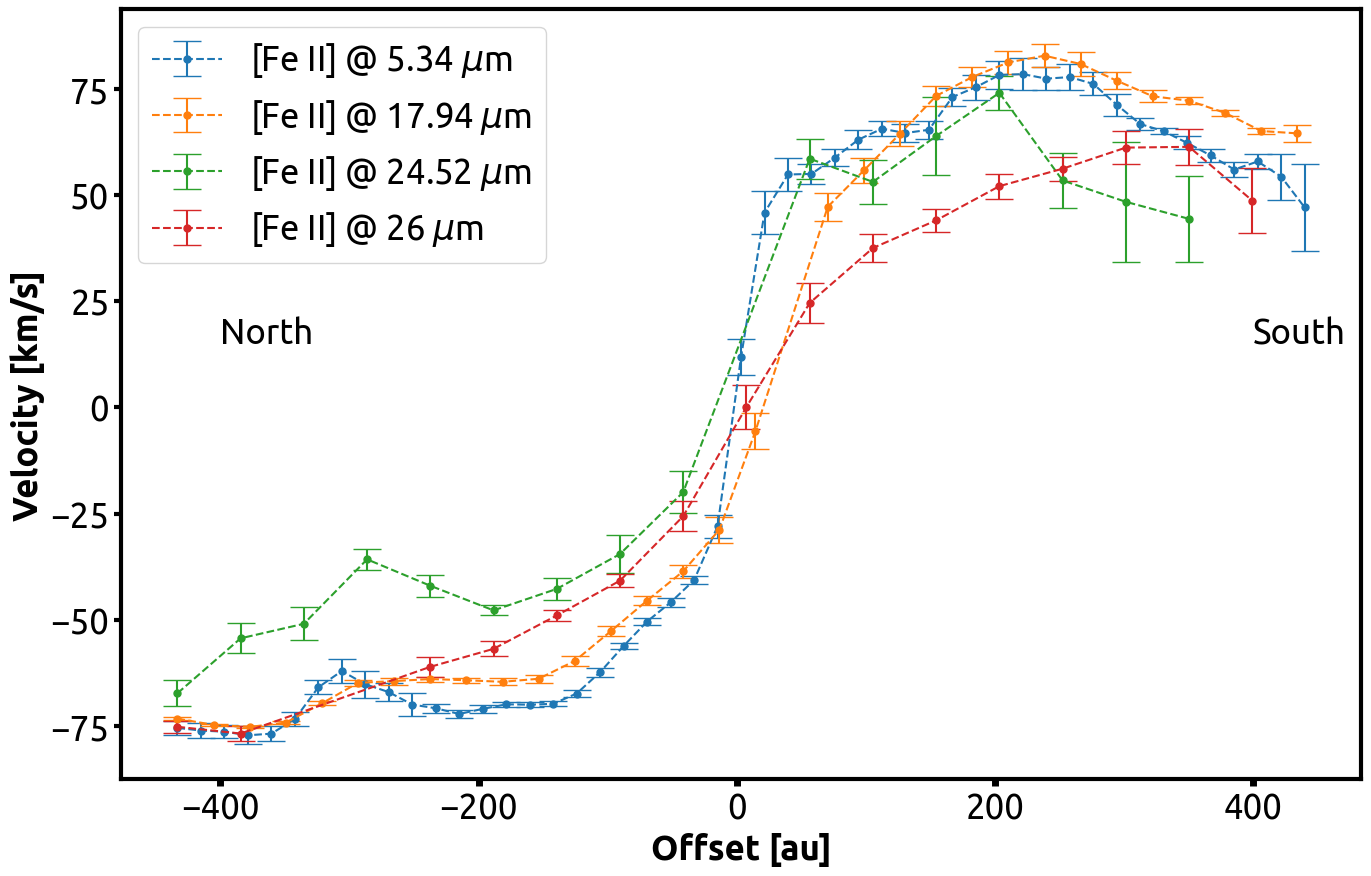}
\caption{The PV diagrams of the [Fe~II] lines {detected in MIRI MRS} aligned to a common average velocity.}
\label{Fig8}
\end{figure*}

\subsection{Velocity structure in the jet}

In order to explore the velocity structures within the jet, we first constructed a moment~1 map of the [Fe~II] line at 5.34~$\mu$m as shown in Figure~\ref{Fig6}. The moment~1 map gives us the intensity-weighted velocity and is an excellent diagnostic tool for understanding the kinematic structures within the jets. As can be seen from Figure~\ref{Fig6}, the northern part of the jet is blue-shifted with respect to the average velocity, while the southern jet is red-shifted. We note that the velocity of the jet on both sides is fairly constant and the maximum blueshift and redshift velocity difference is  $\sim 130$~km\,s$^{-1}$. This is much larger than the systemic cloud velocity (assumed to be V$_\mathrm{LSR}$) of $\sim$4~km\,s$^{-1}$ \citep{2023ApJ...954..101A}.


To further investigate the velocity structure of the jet, we constructed a Position-Velocity (PV) diagram, specifically for the [Fe~II] line at 5.34~$\mu$m. In order to capture the full range of velocities, we selected a rectangular region around the protostar that is both wide and long enough to encompass the entire jet. We show the PV diagram for the [Fe~II] line at 5.34~$\mu$m in Figure~\ref{Fig7}. From the PV diagram, it is apparent that as we move away from the central protostar (the driving source of the jet), the velocities on either side of the jet remain more or less constant. This is similar to what we found with the moment~1 map of [Fe~II] at 5.34~$\mu$m (see Figure \ref{Fig6}) as well. However, within the inner region covering approximately 50~au, we witness rapid changes in velocity {in part due to the blending of emission from the jet and counter-jet}.

{We also constructed PV diagrams for the other [Fe II] lines that were detected in MIRI MRS. We only make PV diagrams of lines detected in MIRI because the NIRSpec IFU has a much lower spectral resolution of $\sim$ 300 km/s \footnote{https://jwst-docs.stsci.edu/jwst-near-infrared-spectrograph}.}   {We computed the velocity of each slice of the spectral cube by converting the wavelength shift with respect to the lab wavelengths \citep{2016JKAS...49..109K, 2018PhRvA..98a2706T} to a velocity shift. In Figure~\ref{Fig8}, we show PV diagrams of all the [Fe~II] lines that we have detected from IRAS 16253$-$2429. } Initially, the PV diagrams of the various [Fe~II] lines did not align since uncertainties in the wavelength calibration can introduce offsets of as large as 40~km\,s$^{-1}$, even within a single channel \citep[especially for wavelengths longwards of 17.7 $\mu$m, channel-3 Long(C)][]{2023A&A...675A.111A}. In Figure \ref{Fig8} we align the velocities in the PV diagrams by setting the mean velocity of the jet seen in each line to zero. Figure \ref{Fig8} shows that the jets display approximately symmetric velocities with respect to the central protostar.  The difference in the peak positive and negative velocities for all the [Fe~II] jets are similar, with an average value of $148 \pm 12$~km\,s$^{-1}$. This consistency in peak jet velocities seen in different [Fe~II] lines reinforces the notion of a coherent jet structure originating from the central protostar. 

{Using the PV diagrams we compute the line-of-sight jet velocity as half of the average peak-to-peak velocity.} The average peak-to-peak velocity is $148 \pm 12$~km\,s$^{-1}$, which gives the line-of-sight jet velocity as $74 \pm 6$~km\,s$^{-1}$. {After correcting for the inclination of the jet using the measured inclination angle of $64.1 \pm 0.5$\arcdeg{} from \cite{2023ApJ...954..101A}, we obtain the velocity of the jet to be  $169 \pm 15$~km\,s$^{-1}$. }

\begin{figure*}
\centering
 \includegraphics[width=0.53\linewidth]{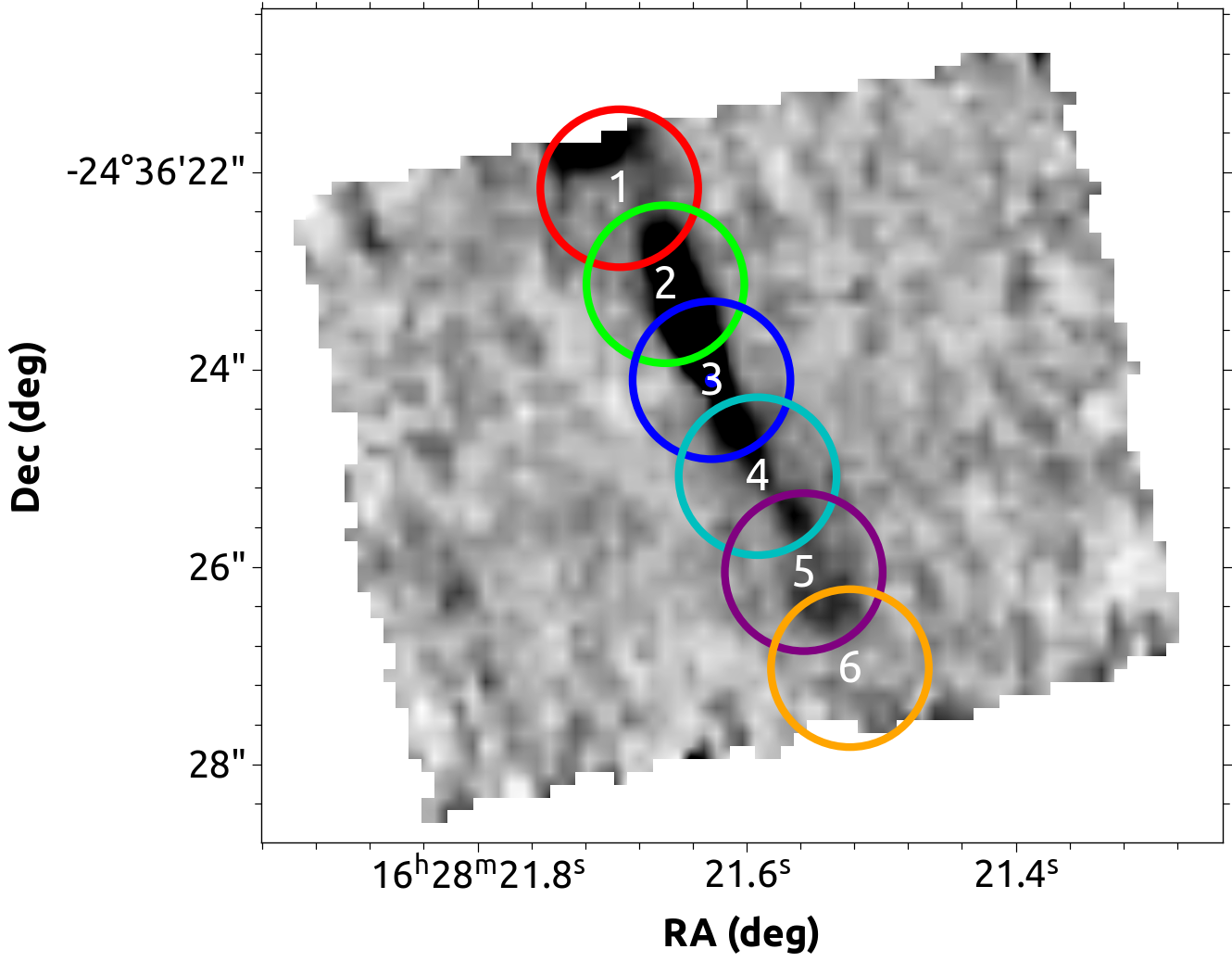}\includegraphics[width=0.5\linewidth]{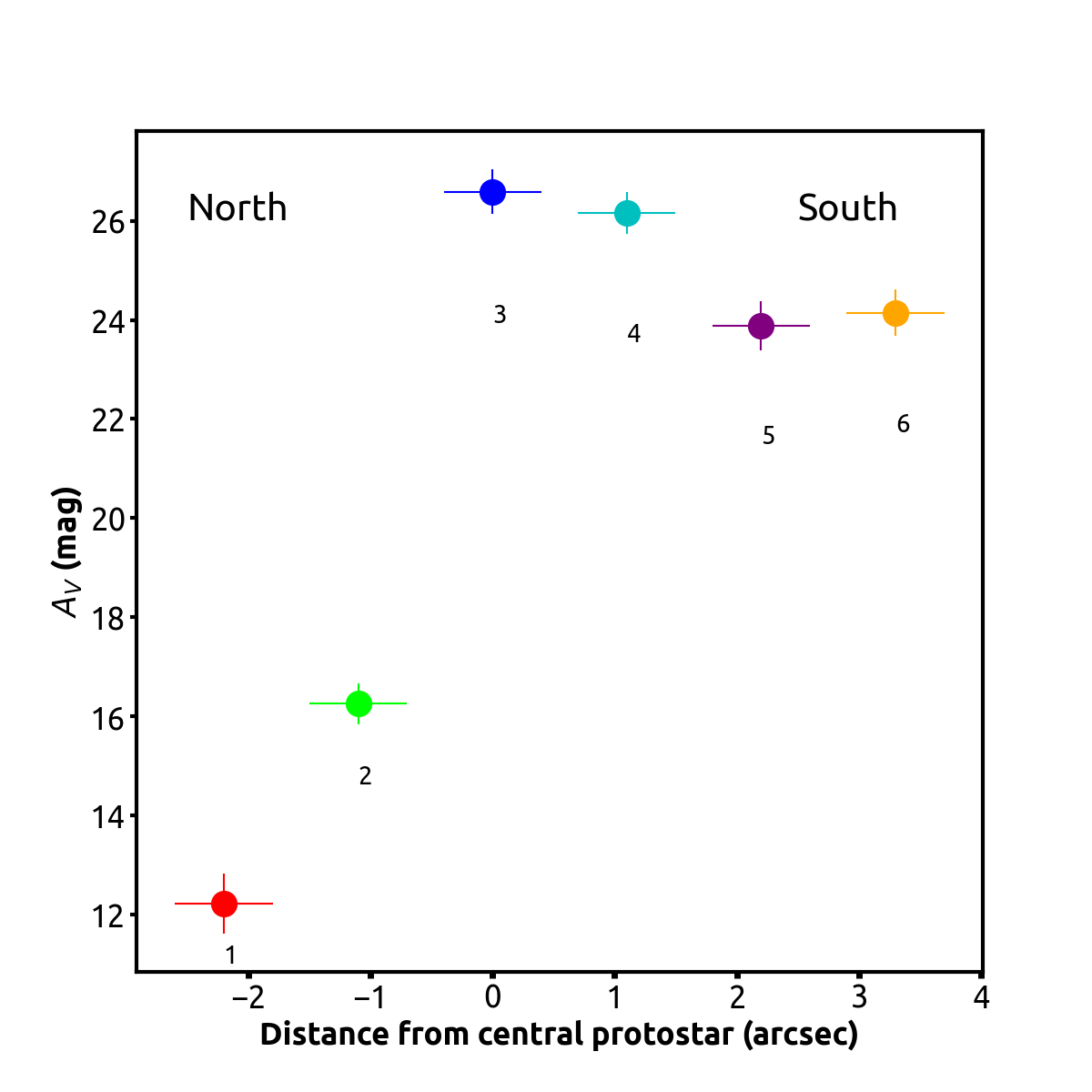}
\caption{(left) The 6 apertures along the jet, where we measure the extinction and jet properties, {overlaid on top of the [Fe II] jet at 5.34 $\mu$m shown in greyscale.} The center of aperture~3 (blue circle) marks the position of the protostar. (right) The measured extinction value along in each of the aperture as a function of their position along the jet {(from the NE to SW)}. }
\label{Fig9}
\end{figure*}

\subsection{Measuring extinction along the jet}

\begin{figure}
\centering
\includegraphics[width=1.1\linewidth]{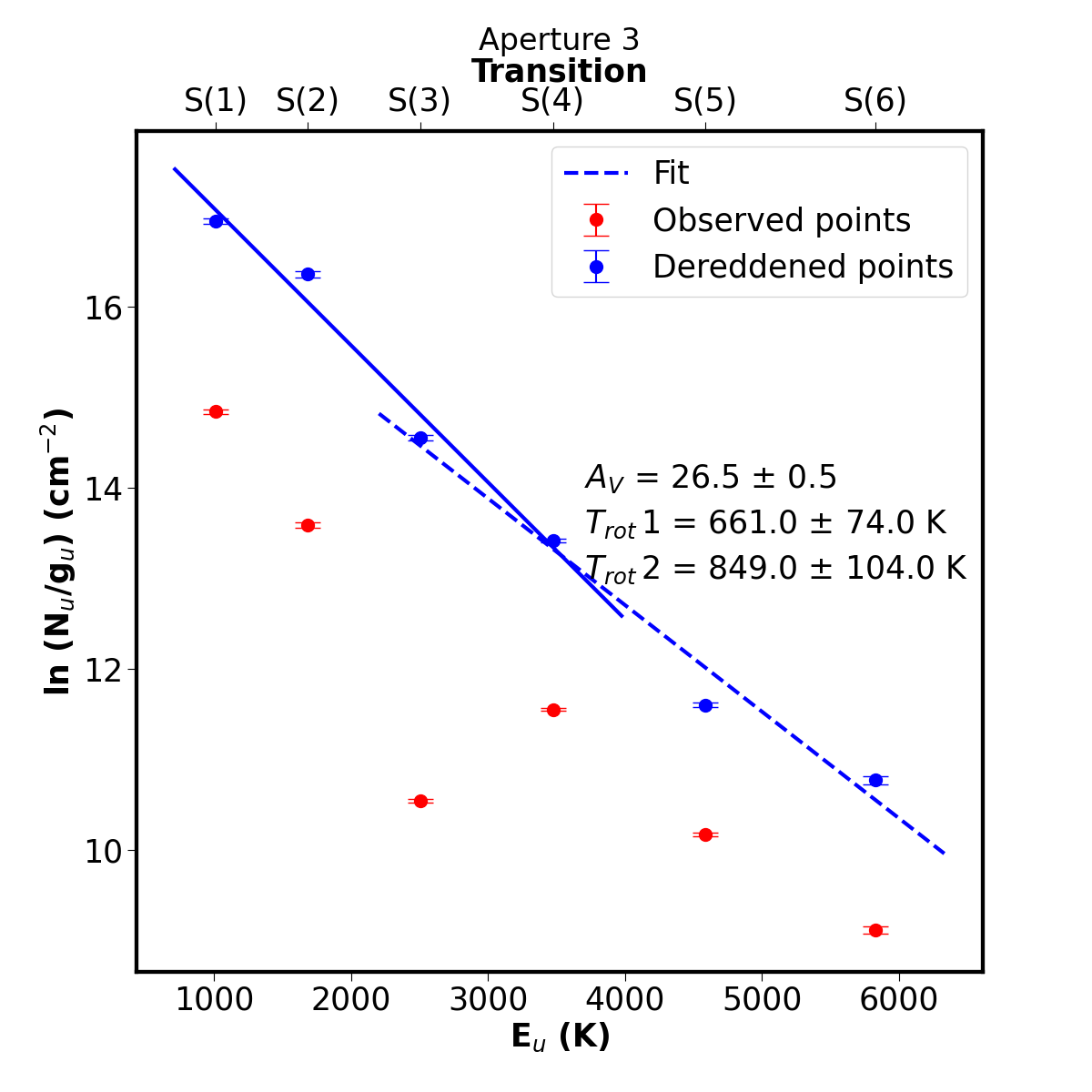}
\caption{The H$_2$ rotation diagram for aperture~3 (centered on the protostar) from Figure~\ref{Fig9}. Also shown are the observed points (red),  the dereddened points (blue)  and a fit to these points. The extinction value $A_{\rm V}$ as well as the rotation temperatures are listed in the figure. {The y-errorbars on the observed (red) points are computed based on the uncertainties in the integrated flux determined by a Gaussian fit to the lines using the measured uncertainties from the IFU error cubes. The y-errorbars on the dereddened (blue) points also incorporate the error in Av from the fit. }}
\label{fig11}
\end{figure}

To derive the physical properties of the jet from the observed line maps, we first measure and quantify the line of sight extinction. To determine the spatially varying extinction towards IRAS 16253$-$2429 outflow cavities, we used the molecular H$_2$~0-0~S(J) lines. Molecular H$_2$ lines have previously been used with Spitzer IRS to determine the extinction values \citep[e.g.,][]{2010ApJ...720...64B}. { Assuming} that the  H$_2$ is in local thermodynamical equilibrium (LTE) with an otho-para ratio of 3 we construct the rotation diagram for the H$_2$ lines (Figure \ref{fig11} and Figure \ref{figA1}). 

In this diagram, the natural logarithm of the column density of H$_2$ molecules in the $u^{ \rm th}$ upper rotational state, $N_u$, divided by the degeneracy of that state $g_u$, $\ln (N_u/g_u)$, is plotted as
a function of the upper state energy, $E_u$ (energy of the $u^{\rm th}$ rotational state) \citep[see, e.g.,][]{2009ApJ...706..170N,2009ApJ...698.1244M,2010ApJ...720...64B, M13}. {Here $g_u = g_{Ju} \times g_s$, with $g_{Ju} = 2Ju+1$ and $g_{s} = 2s+1$, where s = 1 for ortho-H$_2$ and s=0 for para-H$_2$. } 

Assuming that H$_2$ emission is optically thin, then the number of H$_2$ molecules in the $u^{\rm th}$ rotational state is given by

\begin{equation}
    N_u = \frac{4~\pi~F_{u,l}~\lambda_{u,l}}{h~c~A_{u,l}},
\end{equation}

\noindent
where $h$, $c$, $F_{u,l}$, $\lambda_{u,l}$ and $A_{u,l}$ are the Planck's constant, speed of light in vacuum, line flux within the aperture, wavelength, and Einstein A-coefficient, respectively, corresponding to the transition from the upper state $u$ to the lower state $l$. We can compute an average rotational temperature $T_{\rm rot}$, for a relatively small range in $E_u$ by fitting a straight line to the points in the observed rotational diagram. The negative of the inverse of the slope gives us the $T_{\rm rot}$.

Under the LTE assumption, any deviation from this straight line in the narrow range of $E_u$ is due to the extinction. By correcting for these deviations, we derive the line of sight extinction towards the protostars. To do so, we can simultaneously fit the rotation diagram and the correction for extinction. We restrict our analysis to H$_2$ lines between 6-18~$\mu$m (from H$_2$~0-0~S(1) to S(6)) to minimize non-LTE effects and also to avoid contamination from {molecular gas phase CO (also see \citealt{2023arXiv231207807R})}. {In Table 3, we have listed the wavelength, Einstein $A$ coefficient $A_{ul}$, and upper state energy $E_{up}$ of the H$_2$ transitions that are used to compute the line of sight extinction. }

We used the extinction law from Klaus Pontoppidan (hereafter KP v05; private communication) presented in \citep[][]{2009ApJ...690..496C} to derive the extinction towards the protostar (also see section 3.5).  We select apertures of radius 0.\arcsec8 spaced 1.\arcsec1 apart along the jet (see Figure~\ref{Fig9} left) and measure the line fluxes within them.  We then fit the extinction value $A_{\rm V}$ and two rotation temperatures simultaneously, one temperature representing cooler gas between H$_2$~0-0~S(1) and S(4) and the other slightly hotter gas between  H$_2$~0-0~S(3) and S(6)). {In Figure \ref{fig11} we show the  H$_2$ rotation diagram for aperture~3 (centered on the protostar). Correcting for extinction, especially for the H$_2$~S(3) line (as it lies deep within the 10 $\mu$m silicate absorption feature), leads to an improved fit to the rotation diagram. Figure \ref{figA1} in the Appendix shows the H$_2$ rotation diagrams for all apertures. The rotation temperature and $A_{\rm V}$ values for different apertures are listed in Table \ref{Table1}.   }

In Figure \ref{Fig9}(right), we show the variation of the line of sight extinction along the length of the jet. We find that the extinction rapidly rises as we move towards the protostar from North to South, reaching a maximum at the protostellar position, after which it remains relatively constant. A more detailed H$_2$ excitation analysis will be provided in a follow-up paper. 

\begin{figure}
\centering
\includegraphics[width=1.1\linewidth]{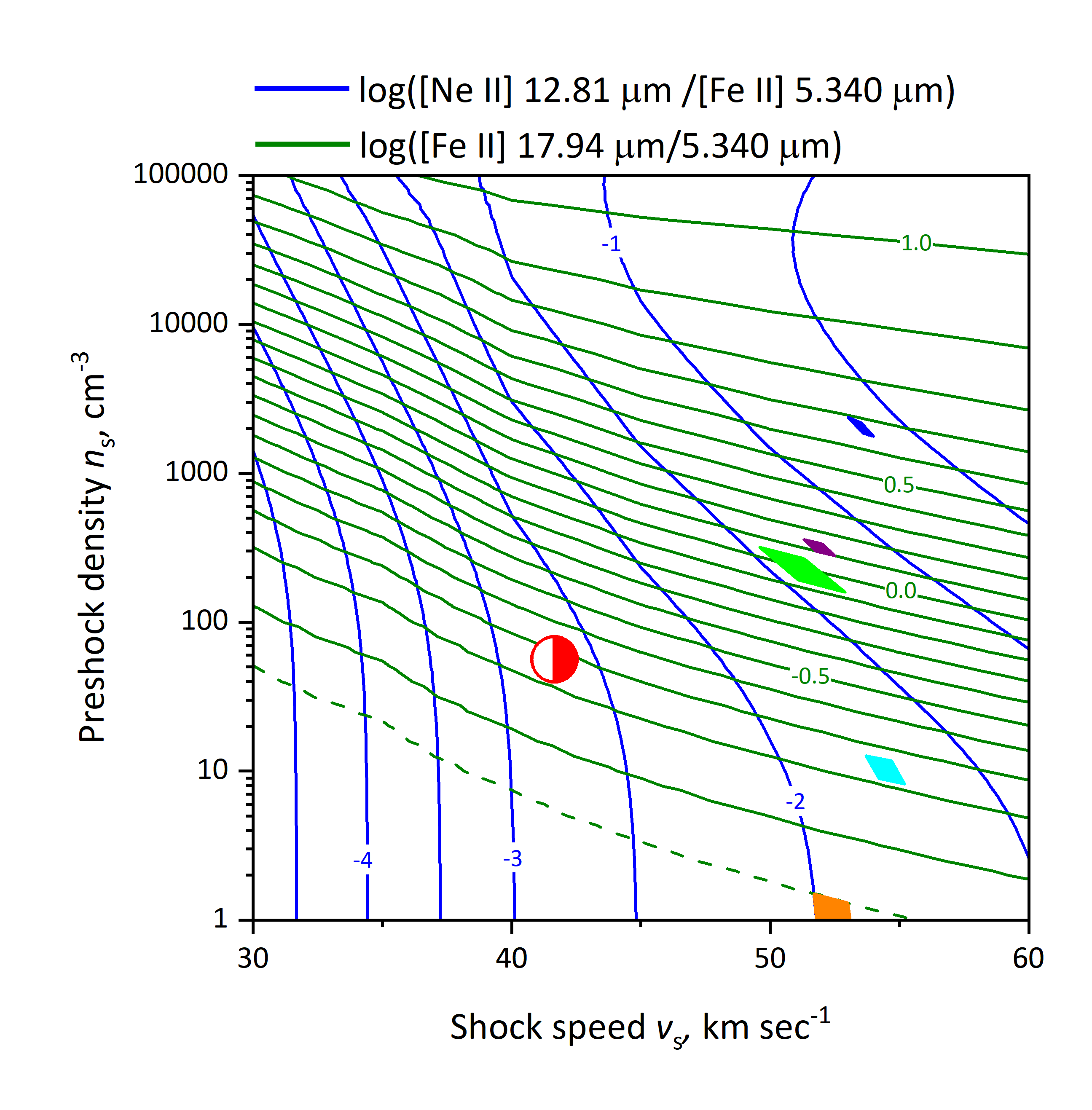}
\caption{Nomogram from the MAPPINGS shock model grid showing the line ratios of three [Fe~II] and [Ne~II] lines. The ratio of [Ne~II] at 12.81~$\mu$m and [Fe~II] at 5.34~$\mu$m in blue runs along the y-axis and is sensitive to the shock {speed}, while the ratio of [Fe~II] lines at 17.93 and 5.34~$\mu$m in green that almost runs parallel to the x-axis is sensitive to the preshock density. {The dashed green curve is the approximate position of the low-density limit to the [Fe II] flux ratio, below which this ratio is insensitive to preshock density.} {The colored regions represent the dereddened line ratios from our observations with uncertainties.  The colors correspond to those of the apertures in  Figure \ref{Fig9}. The red-half-shaded circle represents aperture 1, where the northern knot is not entirely covered by the MIRI channel-1 FOV that has the [Fe II] 5.34~$\mu$m line, resulting in overestimated flux ratios.} }
\label{fig10}
\end{figure}

\subsection{Physical conditions along the jet}

We next investigated the physical conditions (shock {speed} and density) along the jet.  We used the 1D shock and photoionization code MAPPINGS, version 5.1.18 \citep{2018ascl.soft07005S}, to model the shock. Furthermore, we used the [Fe~II] collisional excitation network from \citep{2018PhRvA..98a2706T} instead of the MAPPINGS-default database of atomic properties, {CHIANTI \citep{2021ApJ...909...38D}.} Our shock models were computed for solar abundance \citep{2009ARA&A..47..481A} with a frozen-in magnetic field $B= 15.8 \, \mu {\rm G} \times \sqrt{n/1000\, {\rm cm}^{-3}}$. {We implement the full magnetic and UV radiative precursor treatment by iterating and integrating the shock model to convergence.}

We chose to use [Fe~II] and [Ne~II] as these lines have the largest extent in the jet.  In Figure \ref{fig10} we have plotted a nomogram for these lines from our the MAPPINGS shock model grid. The predicted ratio of [Ne~II] at 12.81~$\mu$m and [Fe~II] at 5.34~$\mu$m shown as blue lines runs along the y-axis and is sensitive to the shock {speed} ($v_{\rm s}$), while the predicted ratio of [Fe~II] lines at 17.93 and 5.34~$\mu$m  shown as green lines (in Figure \ref{fig10}) that almost runs along the x-axis is sensitive to the preshock density ($n_{\rm s}$). By placing the observed line ratios onto this nomogram we can directly read off the shock {speed} and the preshock density.  

The line fluxes along the jet were measured using the same aperture as in Figure \ref{Fig9} (left). The dereddened flux ratios for the [Fe~II] and [Ne~II] lines were then placed onto the grid in Figure \ref{fig10} as colored symbols (the colors scheme for the apertures are the same as those in Figure \ref{Fig9} left) and the shock {speed} and preshock density were read off. We have listed the shock {speed} and preshock density derived for all the apertures in Table \ref{Table1}. Away from the central protostar, the preshock density rapidly falls, while the shock {speed} remains roughly constant.

{The mass loss rate from the protostar, $\dot{M}_{\rm loss}$ can be expressed in terms of the pre-shock density, {speed} of the gas entering the shock, and jet radius as:}

\begin{equation}
    \dot{M}_{\rm loss} = \pi\times R^2 \times n_{\rm s} v \times m_\mathrm{H}
    \label{eq2}
\end{equation}

\noindent
{where  $R$ is the radius of the jet,  $\rho_\mathrm{s}$ is the pre-shock density, $v$ is the velocity of the gas entering the shock, and $m_\mathrm{H}$ is the mass of the hydrogen atom. We have already measured the jet velocity ($v_{\rm jet}$) from the PV diagrams. Based on different jet models, the velocity of gas $v$ used in equation \ref{eq2} can either be the bulk velocity of the jet as derived from the PV diagram  \citep[$v=v_{\rm jet}$; e.g., pulsating jet models of ][]{Raga1, Raga2} or the sum of the shock {speed} and the jet velocity \citep[ $v=v_{\rm jet} + v_{\rm s}$ e.g.,][]{2023ApJ...948...39R}}.

Using the pre-shock density and shock {speed} derived from the shock grid as well as the jet velocity derived from the PV diagram, and assuming that the jet has a cylindrical cross-section with a radius equal to half the jet width measured at the location, we have computed the mass loss rate along the jet from the protostar. The mass loss rates for the various apertures are listed in Table~\ref{Table1}. {From all the models and positions, the highest mass loss rate ($\dot{M}_{\rm loss}$) from the protostar is $1.1 \pm 0.3 \times 10^{-10} M_\odot$/yr.}

In computing $\dot{M}_{\rm loss}$ from the protostar, the largest uncertainty is not due to any measurement uncertainties, but due to the uncertainty in the adopted extinction law.  \cite{2009ApJ...690..496C}  have shown that the extinction law measured towards molecular clouds is highly variable and not only depends upon the molecular cloud but also the environment. To quantify the effects of using different extinction laws on our results, we repeated our analysis of computing extinction and estimating the mass loss rate using two additional extinction laws, that of \cite{2009ApJ...693L..81M} and \cite{2021ApJ...906...73H}. We compute the mass loss rate from the central protostar at the base of the jet (from aperture~3) assuming $v=v_{\rm jet}+v_{\rm s}$. {We find that the mass loss rate obtained using the extinction law from \cite{2009ApJ...693L..81M} is $5.3 \pm 1 \times10^{-11}~M_{\odot}$\,yr$^{-1}$ and using \cite{2021ApJ...906...73H} is  4.1   $\pm 0.8\times10^{-11}~M_{\odot}$\,yr$^{-1}$. These $\dot{M}_{\rm loss}$ values and the ones listed in Table~\ref{Table1} (for aperture~3) are within a factor of 3, indicating that the mass loss rate of $\dot{M}_{\rm loss}~ 1.1 \pm 0.3 \times10^{-10}~M_\odot$\,yr$^{-1}$ is robust}. 

\section{Discussion}
Our JWST observations reveal a highly collimated atomic jet in the IR fine structure lines and H~I lines from the low luminosity and very low mass protostar IRAS~16253$-$2429. No collimated molecular jet is detected, 
and although IR H$_2$ emission is detected along the outflow cavity, it has a much wider opening angle and does not spatially follow the collimated atomic jet. The jet seen in the FS lines appears brighter on the blueshifted north-eastern side that is facing us, and is fainter to the south-west; this is due to the higher line-of-sight extinction toward the south-western cavity (see Figure \ref{Fig9}). 

The jets on either side appear to have similar velocities that more or less remain constant as a function of distance from the driving source. The jet velocity, corrected for inclination effects (using the inclination of 64.1\arcdeg), is $169 \pm 15$~km\,s$^{-1}$. The collimated jet seen in the JWST line images extends up to about 400~au (see Figure~\ref{Fig8}; 445~au after correcting for inclination) from the driving source on either side. This implies a dynamical timescale of $\sim$12.5~yr, suggesting that the jet seen in the JWST IFUs was ejected only in the last 12.5~yr. 

{The width of the jet (FWHM) at the base is marginally resolved and is $\sim23 \pm 4$~au, and the jet tends to widen as it moves away from the central source with an opening angle $\Theta = 2.6$\arcdeg. If the jet is ballistically confined, then the opening angle can be expressed as a ratio of the sound speed within the jet and the jet speed.}


\startlongtable 
\begin{deluxetable*}{cllllcc}
    \tablewidth{0pt}
    \tablecaption{The H$_2$ lines used to estimate the extinction along the jet. The wavelength, Einstein A coefficient $A_{ul}$, and upper state energy are from \cite{HITRAN}.  }
    \tablehead{
    \colhead{Wavelength} &  \colhead{Name} &  \colhead{Upper}  &  \colhead{Lower} & \colhead{$A_{ul}$} & \colhead{$E_{up}$}  \\
    \colhead{$\micron$} & \colhead{} & \colhead{state} & \colhead{state} &    \colhead{s$^{-1}$}   &  \colhead{K} \\
    }
    \decimals
    \startdata
17.035	 & 	H$_2$ 0-0 S(1)	 & 	v = 0, J=(3)  	 & 		v = 0, J=(1)	 		 & 	4.758 $\times 10^{-10}$ 	 & 		1015		 \\
12.279	 & 	H$_2$ 0-0 S(2)	 & 		v = 0, J=(4)   & 		v = 0, J=(2)	 		 & 	  2.753 $\times  10^{-9}$	 & 		1681		 \\
9.665	 & 	H$_2$ 0-0 S(3)	 & 	v = 0, J=(5) 	 & 		v = 0, J=(3)	  		 & 	 9.826 $\times 10^{-9}$ 	 & 		2503		 \\
8.025	 & 	H$_2$ 0-0 S(4)	 & 		v = 0, J=(6)	 & 		v = 0, J=(4)	  	 & 	 2.641 $\times 10^{-8}$ 	 & 		3474		 \\ 
6.910	 & 	H$_2$ 0-0 S(5)	 & 	v = 0, J=(7) 	 & 		v = 0, J=(5)	  		 & 	 5.872 $\times 10^{-8}$ 	 & 		4586		 \\
6.109	 & 	H$_2$ 0-0 S(6)	 & 	v = 0, J=(8)	 & 		v = 0, J=(6)	 		 & 	 1.140 $\times 10^{-7}$ 	 & 		5830	 \\ \enddata

      \vspace{-0.3in}
           \label{TableH2}
\end{deluxetable*}

\begin{equation}
 \Theta = \frac{c_{\rm s}}{v_{\mathrm{jet}}},
\end{equation}

\noindent
where $v_{\mathrm{jet}}$ is the jet velocity of $169 \pm 15$~km\,s$^{-1}$ and $c_{\rm s}$ is the sound speed \citep{2021NewAR..9301615R}. If we assume that the shock temperature is 10,000~K \citep[e.g.,][]{2021NewAR..9301615R} then the {sound speed $c_{\rm s} = 10$~km\,s$^{-1}$, which gives an opening angle of $3.4 \pm 0.3$\arcdeg. This is similar to the opening angle we measure from the observed jet widths, given the uncertainties. This suggests that the jet is likely ballistically confined, at least within the central 40-1000~au of the driving source; other confinement mechanisms, however, could be at play at larger distances.} 

The physical conditions of the jet derived from our analysis indicate shock {speed} of 42-54~km\,s$^{-1}$ from the base to one end of the jet, suggesting that there is no significant acceleration along the jet. The preshock densities estimated are in the range of $\sim$10-2000~cm$^{-3}$, indicating that the jet is rather tenuous.

\begin{sidewaystable*}[]
\centering
\caption{The extinction $A_{\rm V}$, the rotation temperature of the H$_2$ gas (T$_{\rm rot}$1 and T$_{\rm rot}$2), the shock {speed}, preshock densities,  jet radius $R$ (computed using the opening angle $\Theta$ and the jet width at the base) along with the mass loss rate $\dot{M}_{\rm loss}$ for each of the apertures shown in Figure \ref{Fig9}. {The errors associated with the mass loss rate incorporate the uncertainties of the shock speed and pre-shock density, which are determined from the uncertainties in the line flux and the extinction  $A_{\rm V}$, the uncertainty in the measurement of the jet width and that of the $v_{\mathrm{jet}}$ ($v_{\mathrm{jet}}$ = $169 \pm 15$~km\,s$^{-1}$).}   {{$^*$ In aperture 1 the northern knot is not entirely covered by the MIRI channel-1 FOV therefore the flux ratio might be overestimated.} } }
\begin{tabular}{|c|c|c|c|c|c|c|c|c|c|}
\hline
Aperture & Distance from & $A_{\rm V}$& T$_{\rm rot}$1 & T$_{\rm rot}$2 &  $v_{\rm s}$  & $n_{\rm s}$& $R$ & $\dot{M}_{\rm loss}$ for & $\dot{M}_{\rm loss}$ for\\
  & Protostar  & & &  &  &  & & $v=v_{\rm jet}+v_{\rm s}$ & $v=v_{\rm jet}$  \\

 & (\arcsec) & &  (K) & (K)   & (km\,s$^{-1}$) & (cm$^{-3}$)& (au)  &($M_{\odot}$~yr$^{-1}$) & ($M_{\odot}$~yr$^{-1}$) \\ \hline
1$^*$ & -2.2 & 12.3 $\pm$ 1.1 &526 $\pm$ 35 & 797 $\pm$ 75 & (42)  & (6$\times 10^{1}$)& 25.3 $\pm$ 5.1 &1.5 $\times10^{-11}$ & 1.2$\times10^{-11}$ \\

2 & -1.1 & 16.3 $\pm$0.7 &585 $\pm$ 42 & 870 $\pm$ 88 & 51 $\pm$ 2 & 2.4 $_{+0.8}^{-0.8}$ $\times 10^{2}$ &18.4 $\pm$ 3.6 &3.3 $\pm$ 1.3 $\times10^{-11}$ &2.6 $\pm$1 $\times10^{-11}$\\

3 & 0 & 26.6 $\pm$ 0.5 &661 $\pm$ 74 & 849 $\pm$ 104  & 54 $\pm$ 1 &  2 $_{-0.3}^{+0.4}$ $\times 10^{3}$& 11.5 $\pm$ 2 & 1.1 $\pm$ 0.3 $\times10^{-10}$ &8 $\pm$ 2 $\times10^{-11}$\\
4 & 1.1 & 26.1 $\pm$ 0.6 &685 $\pm$ 63 & 852 $\pm$ 101  & 52 $\pm$ 1 &  3.2 $_{-0.4}^{+0.4}$ $\times 10^{2}$&18.4 $\pm$ 3.6 &4.5 $\pm$ 1$\times10^{-11}$ &3.4 $\pm$ 0.8 $\times10^{-11}$\\

5 & 2.2 & 23.9 $\pm$ 0.6 &621 $\pm$ 54 & 840 $\pm$ 96  & 54 $\pm$ 1 & 1$_{-0.3}^{+0.2}$ $\times 10^{1}$&25.3 $\pm$ 5.1 &2.7 $\pm$ 0.9 $\times10^{-12}$ &2 $\pm $0.7 $\times10^{-12}$\\

6 & 3.3 & 24.2 $\pm$0.7 &672 $\pm$ 46 & 883 $\pm$ 101  & 52 $\pm$ 1 & $<$ 2 &32.3 $\pm$ 6.6 &$<$8.6$\times10^{-13}$ &$<$6.6$\times10^{-13}$\\ \hline
\end{tabular}%

\label{Table1}
\end{sidewaystable*}

{The mass loss rate $\dot{M}_{\rm loss}$ measured from the central protostar as calculated for a suite of jet models and extinction laws is between $ 0.4 - 1.1 \times 10^{-10} ~M_{\odot}$~yr$^{-1}$.}  The estimated mass loss rate is low, even for a very low mass protostar. { The simultaneous accretion rate derived from the OH lines in the MIRI spectrum of IRAS~16253$-$2429 is $\dot{M}_{\rm acc}~=2.4~\pm~0.8~\times~10^{-9}~M_{\odot}$~yr$^{-1}$ (Watson et al. in prep), is commensurate with the low mass loss rate observed. The protostar appears to be going through a phase of low mass accretion/ejection currently. It is, however, driving a highly collimated jet, even while in the relatively quiet phase, suggesting that collimated jets are present even when protostars are passing through a low accretion phase \citep[also see][]{2023ApJ...954..101A}.}

{Although the protostar IRAS~16253$-$2429 is currently in a quiescent phase, it must have gone through earlier epochs of high mass accretion/ejection, and is also very likely to go through active high accretion phases later in its life, for the following reasons.}

{The current mass of the central source based on the ALMA observations of the Keplerian disk is about 0.15~$M_\odot$ \citep{2023ApJ...954..101A}. At the current mass accretion rate, it would take more than 100~Myr to accrete a mass of 0.15~$M_\odot$. IRAS 16253$-$2429 is a Class~0 protostar, which is likely to be $\ll 1$~Myr old \citep[e.g.,][]{2014prpl.conf..195D}, suggesting that the protostar must have had a period(s) of much higher accretion rate at an earlier stage during which most of the current mass was accreted.  Additionally, wide cavities are already present in the envelope of IRAS~16253$-$2429 \citep[see][]{2023arXiv231003803F}, which must have been carved out by powerful jets and outflows associated with an earlier high accretion phase. All these evidences indicate that IRAS~16253$-$2429 must have been accreting at a significantly higher rate at an earlier time in its life. 
 \cite{2022ApJ...924L..23Z} have shown that during these high accretion phases, a protostar can accrete as much as 100\%\ of its current mass.}

The envelope mass of IRAS~16253$-$2429 estimated from mm observations is in the range of 0.2 to 1~$M_\odot$ \citep{2006A&A...447..609S, 2008ApJ...684.1240E, 2011ApJ...740...45T} depending on the size of the core considered, the opacity values used, and the assumed dust temperature. All measurements and our modeling efforts are consistent with a mass of $\gg$~0.2~$M_\odot$ within an envelope radius of $\sim$4200~au, suggesting that there is significantly more mass in the envelope of IRAS~16253$-$2429 than in the central object. {If we assume an efficiency of $\sim$30\% in converting envelope mass to stellar mass \citep{K10,A10}, then the available mass in the envelope (between 0.2 - 1~$M_\odot$) would suggest at least an additional 0.06 - 0.15~$M_\odot$ will be accreted onto the central source. This would make the final mass of IRAS~16253$-$2429 to be $\sim$0.2 - 0.3$~M_\odot$, which is where the stellar initial mass function peaks \citep{2010ARA&A..48..339B}. }

{At the current accretion rate, it would take {$>50$~Myr} to accrete this mass, strongly suggesting that the protostar will again go into a high mass accretion/ejection phase. This is consistent with the results of \cite{2022ApJ...924L..23Z}, who find that Class~0 protostars undergo bursts every few hundred years that last only a few decades; hence it is more likely that a protostar is observed in between bursts.  In the case of one Class~0 protostar, HOPS~383, the pre and post- burst periods are characterized by a quiescent state with a very low accretion rate \citep{2022ApJ...924L..23Z}.  Our JWST observations appear to have captured IRAS 16253$-$2429 during the interlude between the accretion bursts, i.e. in the quiescent phase, and yet driving a tenuous but highly collimated jet.}

{The low mass loss rate that we derive for the protostar is also consistent with the non-detection of a molecular jet from the source. \cite{2020A&A...636A..60T} have argued that self-shielding of the jet from FUV photons generated from the central source or from shocks along the jet requires much higher mass loss rates from the protostar.  }

\section{Summary and Conclusions}
We report the discovery of a  highly collimated {atomic} jet as detected in multiple [Fe~II] and other fine structure lines and in H~I, originating from the Class~0 protostar IRAS 16253$-$2429 in the Ophiuchus molecular cloud (140 pc), the least luminous source in the "Investigating Protostellar Accretion" (IPA) program.  No collimated molecular jet is detected in CO, SiO, in the (sub-)mm with ALMA or H$_2$ in the IR with JWST. We determine the following properties for the jets/outflows from IRAS 16253-2429:

\begin{enumerate}
    \item The jet (as seen in the [Fe~II] line at 5.34~$\mu$m) is $23 \pm 4$~au wide at the base with an opening angle of $2.6 \pm 0.5$\arcdeg.

    \item The jet {seen in [Fe II] lines} has a velocity of $169 \pm 15$~km\,s$^{-1}$ after correcting for inclination. The jets on either side of the protostar appear to have similar velocities that remain almost constant as a function of distance from the center.

    {\item The jet is about 445~au long (after correcting for inclination) on either side with a dynamical timescale of $\sim$12.5~yr.}

    \item We determine the line of sight extinction  $A_{\rm V}$ along the jet using H$_2$ lines in the cavity, which gives $A_{\mathrm{V}}$ values ranging from 12.3 to 26.6 mag.

    \item Using the extinction corrected flux ratios of [Ne~II] at 12.81~$\mu$m and the [Fe~II] lines at 5.34~$\mu$m and 17.936~$\mu$m, we derive a shock {speed} of 54~km\,s$^{-1}$ and a preshock density of $2.0 \times 10^{3}$~cm$^{-3}$ at the base of the jet. 

    \item  Assuming a cylindrical jet with a cross-sectional width of 23~au and using the jet velocity of 169~km\,s$^{-1}$ along with the derived preshock density and shock {speed}, {we compute the mass loss rate from the jet to be $0.4-1.1 \times10^{-10}~M_{\odot}$~yr$^{-1}$. }
    
\end{enumerate}

 The low mass loss rate derived is consistent with the simultaneous measurements of low accretion rate of \mbox{$\sim 2.4\times 10^{-9}~M_{\odot}$~yr$^{-1}$} for IRAS~16253$-$2429 (Watson et al. in prep), indicating that the protostar is likely in the quiescent accretion phase. Our results show that even a very low mass protostar in a quiescent phase with low accretion rate can still drive a highly collimated jet.

\section{Data availability}
{All of the data presented in this article were obtained from the Mikulski Archive for Space Telescopes (MAST) at the Space Telescope Science Institute. The specific observations analyzed can be accessed via \dataset[DOI: 10.17909/3kky-t040]{https://doi.org/10.17909/3kky-t040}.}

\section{Acknowledgment}
This work is based on observations made with the NASA/ESA/CSA James Webb Space Telescope. The data were obtained from the Mikulski Archive for Space Telescopes at the Space Telescope Science Institute, which is operated by the Association of Universities for Research in Astronomy, Inc., under NASA contract NAS 5-03127 for JWST. These observations are associated with program \#1802. Support for SF, AER, STM, RG, WF, JG, JJT and DW in program \#1802 was provided by NASA through a grant from the Space Telescope Science Institute, which is operated by the Association of Universities for Research in Astronomy, Inc., under NASA contract NAS 5-03127. A.C.G. has been supported by PRIN-MUR 2022 20228JPA3A “The path to star and planet formation in the JWST era (PATH)” and by INAF-GoG 2022 “NIR-dark Accretion Outbursts in Massive Young stellar objects (NAOMY)”. G.A. and M.O., acknowledge financial support from grants PID2020-114461GB-I00 and CEX2021-001131-S, funded by MCIN/AEI/10.13039/501100011033. Y.-L.Y. acknowledges support from Grant-in-Aid from the Ministry of Education, Culture, Sports, Science, and Technology of Japan (20H05845, 20H05844, 22K20389), and a pioneering project in RIKEN (Evolution of Matter in the Universe). {W.R.M.R. thanks support from the European Research Council (ERC) under the European Union’s Horizon 2020 research and innovation programme
(grant agreement No. 101019751 MOLDISK) . {We thank the anonymous referee for their valuable comments. MN would lastly like to thank his Nani for her constant support. }}

\facility{JWST (NIRSpec, MIRI)}

\appendix

\begin{figure}
\centering
\includegraphics[width=0.35\linewidth]{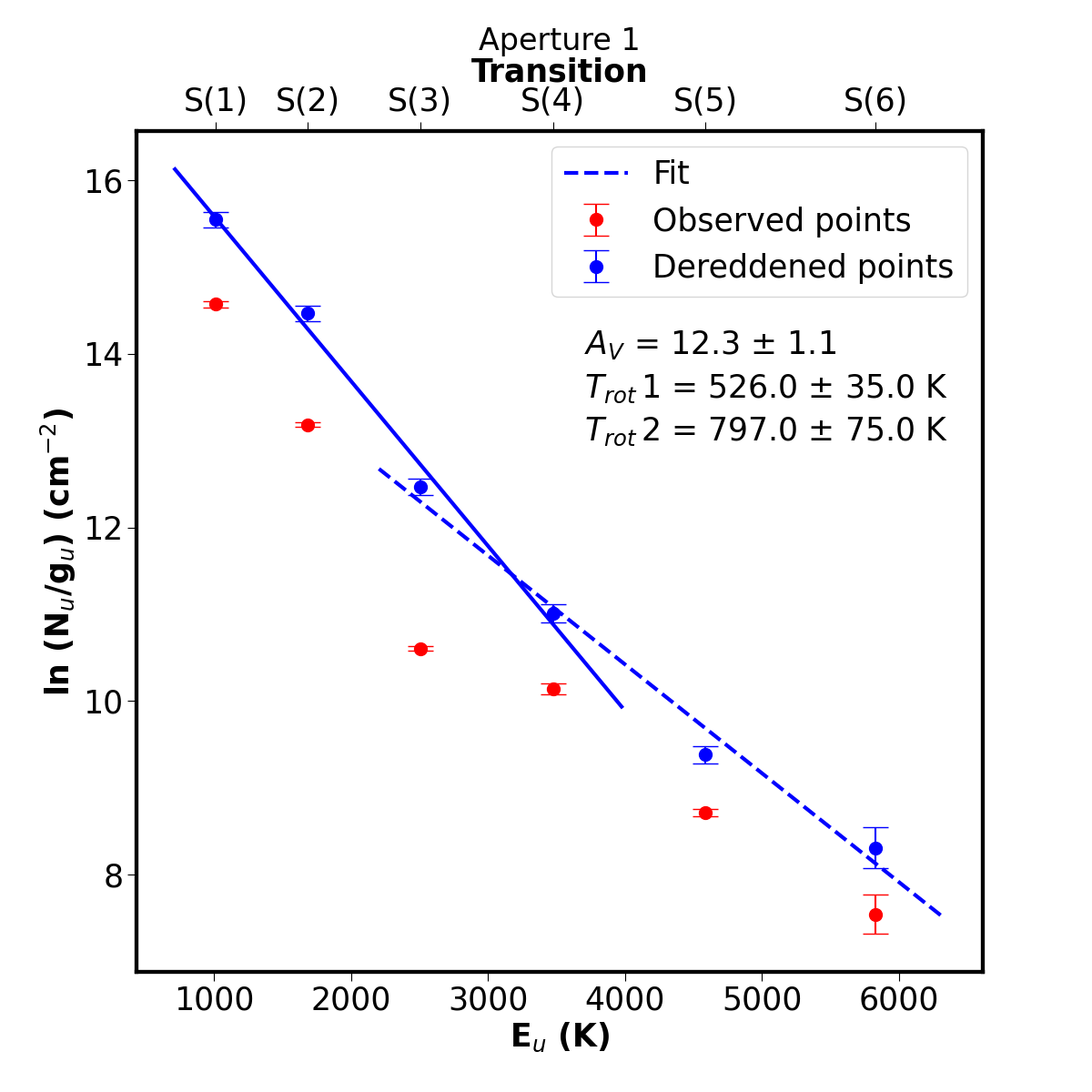}\includegraphics[width=0.35\linewidth]{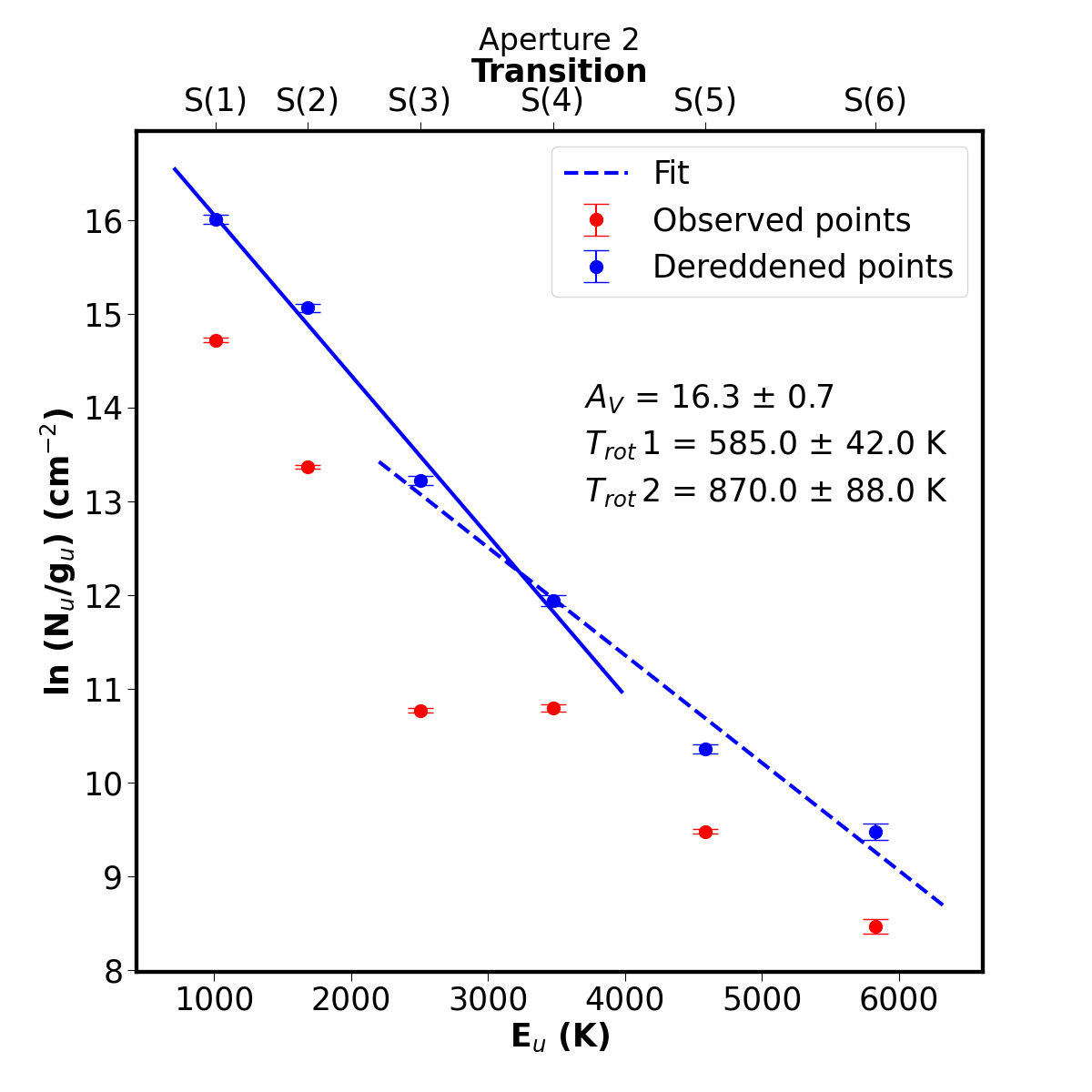}
\includegraphics[width=0.35\linewidth]{rt_rot_KP_Ap=3.png}\includegraphics[width=0.35\linewidth]{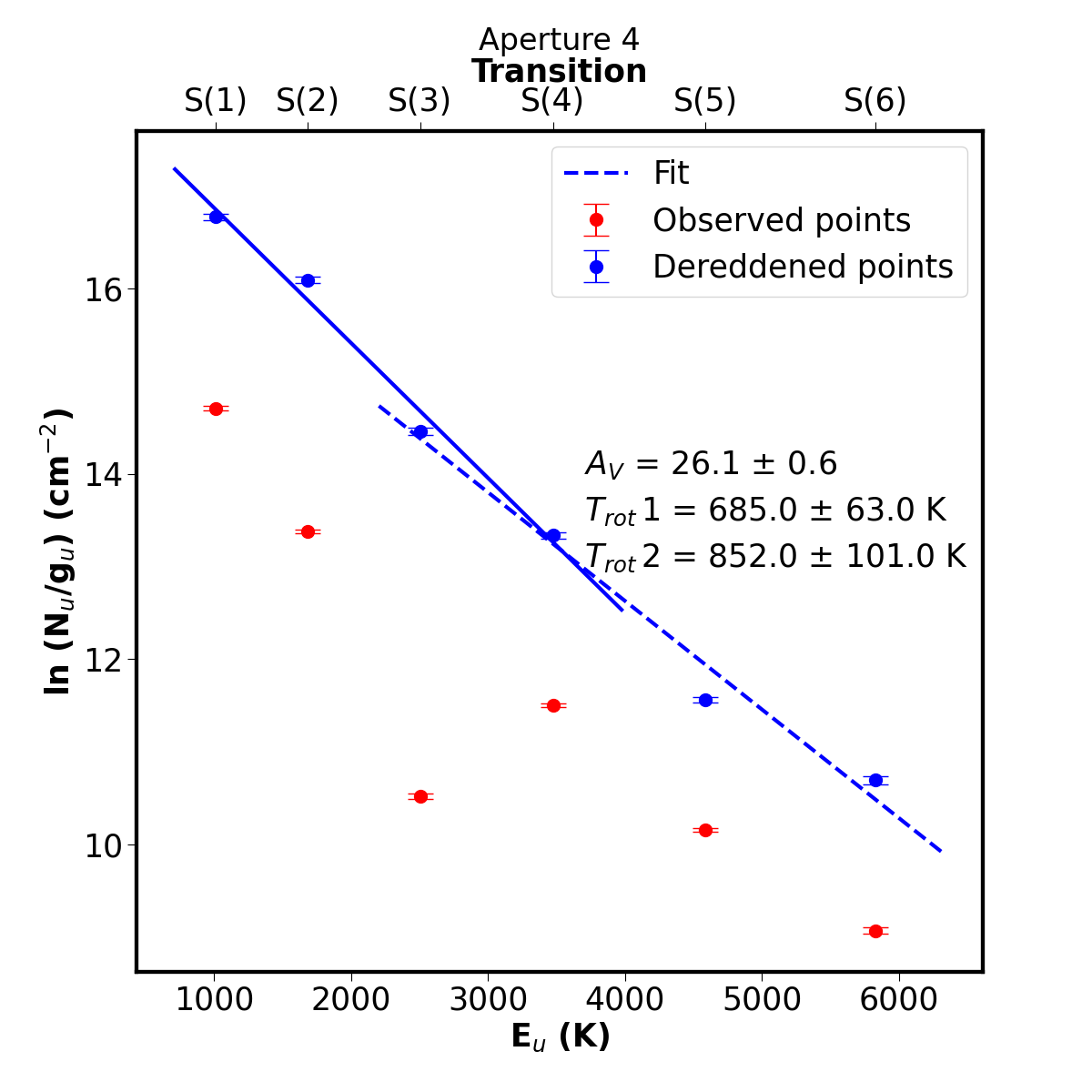}
\includegraphics[width=0.35\linewidth]{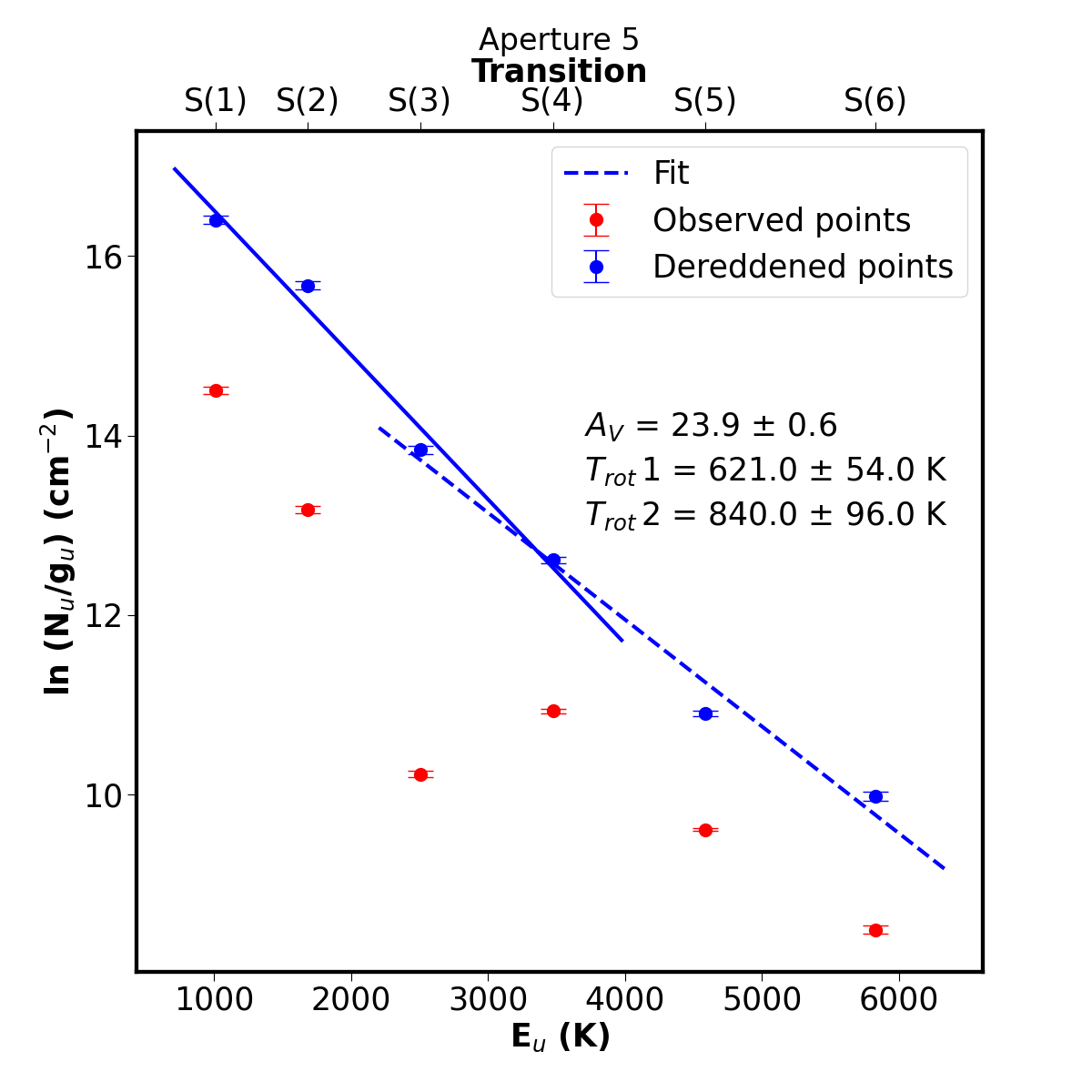}\includegraphics[width=0.35\linewidth]{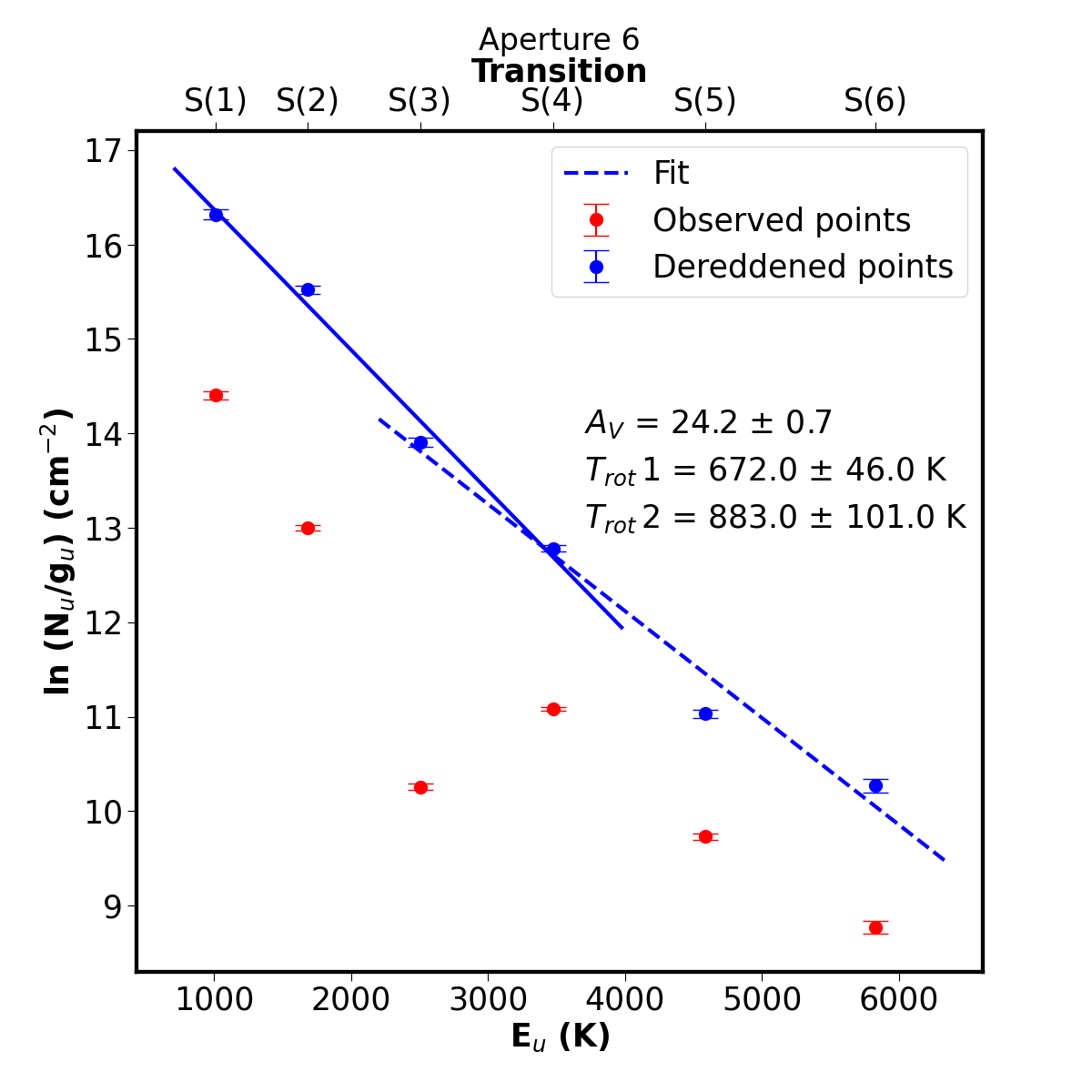}
\caption{{The H$_2$ rotation diagram for all  apertures  shown in Figure \ref{Fig9}. Also shown are the observed points (red), the dereddened points (blue), as well as the well as the fit to them. The extinction values $A_{\rm V}$ as well as the rotation temperatures are listed in the figure. {The y-errorbars on the observed (red) points are computed based on the uncertainties in the integrated flux determined by a Gaussian fit to the lines using the measured uncertainties from the IFU error cubes. The y-errorbars on the dereddened (blue) points also incorporate the error in Av from the fit. }}}
\label{figA1}
\end{figure}

\end{document}